\documentclass[12pt]{article}
\usepackage{graphics, amsfonts, amsmath, mathrsfs}
\def\preprint
#1{\thispagestyle{empty}~\newline\vspace*{-22.65mm}
\begin{flushright}
\begin{tabular}{l} #1
\end{tabular}
\end{flushright}
\vspace{1cm}}
\begin{document}
%--------------------------------------------------------------------
\markright{\hfil Coordinates, observables and symmetry in relativity}
%--------------------------------------------------------------------
\title{\bf \LARGE Coordinates, observables and symmetry in relativity}

\author{Hans Westman~$^*$ and Sebastiano Sonego~$^\dagger$\\[2mm]
%-----------------------------------
{\small\it\thanks{\tt hwestman@physics.usyd.edu.au}
\ Physics Building A28, University of Sydney, NSW 2006}\\
{\small\it Sydney, New South Wales, Australia}\\[4mm]
%-----------------------------------
{\small\it\thanks{\tt sebastiano.sonego@uniud.it}
\ Universit\`a di Udine, Via delle Scienze 208, 33100 Udine,
Italy}}
%--------------------------------------------------------------------
\date{{\small June 30, 2009; \LaTeX-ed \today }}
%--------------------------------------------------------------------
\maketitle
%--------------------------------------------------------------------
\begin{abstract}
We investigate the interplay and connections between symmetry properties of equations, the interpretation of coordinates, the construction of observables, and the existence of physical relativity principles in spacetime theories.  Using the refined notion of an event as a ``point-coincidence'' between scalar fields that completely characterise a spacetime model, we also propose a natural generalisation of the relational local observables that does not require the existence of four everywhere invertible scalar fields.  The collection of all point-coincidences forms in generic situations a four-dimensional manifold, which is naturally identified with the physical spacetime.\\

\noindent PACS: 04.20.Cv, 01.70.+w, 04.90.+e, 03.30.+p\\
Keywords: Spacetime; events; general covariance; observables; hole argument; symmetry; relativity principles
\end{abstract}
%-----------------------------------------------------------------------

\def\g{\mbox{\sl g}}
\def\eg{{\em e.g.\/}}
\def\ie{{\em i.e.\/}}

\newpage
%-----------------------------------------------------------------------
\section{Introduction}
\label{sec:intro}
\setcounter{equation}{0}
%-----------------------------------------------------------------------

In spacetime theories~\cite{friedman}, a model of spacetime is usually a pair $(\mathscr{M},T)$, where $\mathscr{M}$ is a four-dimensional manifold with suitable topological and differentiable properties, and $T$ represents a collection of tensor fields on $\mathscr{M}$.

The modern presentations of differential geometry, in which one starts from a bare set $\mathscr{M}$ and progressively adds structure to it, suggest an interpretation of the manifold $\mathscr{M}$ as an independently existing ``container'' for the histories of fields and particles.  That is, although it is the fields $T$ that, according to this interpretation, represent the geometrical and physical properties at various points of $\mathscr{M}$, such points are taken as existing independently of those properties and of the fields themselves.\footnote{If the set $\mathscr{M}$ (the ``container'') is postulated before any field is defined on it, it is natural to think of $\mathscr{M}$ as having an ontological status of its own.  Considering $\mathscr{M}$ as an independently existing entity, corresponds broadly with the doctrine that philosophers of science call {\em spacetime substantivalism\/}~\cite{earman, pooley}.}

This view, which gives the points of $\mathscr{M}$ an ontological status and purports that the fields on $\mathscr{M}$ somehow ``individuate'' such points, has to be counterposed to another one, in which the spacetime manifold is conceptually identified with the total collection of events constructed out of the actual physical and geometrical fields, so its existence trivially {\em requires\/} the presence of such fields.\footnote{Noteworthy, these two interpretations, although mutually contradictory, sometimes coexist in textbooks on general relativity.} Apparently, this was Einstein's conception, as the following quotations suggest:
\begin{quote}
{\em According to general relativity, the concept of space detached from any physical content does not exist\/}~\cite{einstein50};
\end{quote}
\begin{quote}
{\em There is no such thing as an empty space, i.e., a space without field.  Space-time does not claim existence on its own, but only as a structural property of the field\/}~\cite{einstein};
\end{quote}
\begin{quote}
{\em Physical objects are not in space, although they are spatially extended\/}~\cite{einstein-preface}.
\end{quote}

One purpose of the present paper is to show in detail how such a radical (and perhaps surprising) view emerges naturally within the theory of general relativity.  We shall see how one can construct from the actual physical and geometrical fields a set that, except in pathological situations, forms a four-dimensional manifold and in addition contains all the observable properties of the spacetime model.  We denote this set the {\em space of point-coincidences\/} $\mathscr{E}$ in order to distinguish it from $\mathscr{M}$.  We shall show that spacetime itself can be naturally identified with $\mathscr{E}$, and that $\mathscr{M}$ need not be given any ontological status.  Another purpose is to clarify the notion of observables in general relativity and to propose a new way to think about such quantities.

Our starting point will be the notion of general invariance, that we present in section~\ref{sec:invariance} as a symmetry property of Einstein's field equations.  In section~\ref{sec:hole} we first discuss in some detail the notion of coordinates in special relativity, pointing out how, in such a theory, two interpretations are possible: One in which the coordinates have operational significance; and one in which they are just parameters with no operational meaning.  Then, we show that general invariance forces one to adopt the latter interpretation, unless one wants to accept that general relativity cannot make unique {\em empirical\/} predictions. Dropping the operational significance of the coordinates leads one to wonder, on the one hand, what the observable quantities are in general relativity; and, on the other hand, how spacetime events can be mathematically represented in the theory.  These two problems are deeply intertwined, and will be treated in sections~\ref{sec:observables} and~\ref{sec:events}.  The picture of spacetime that emerges elucidates the previous quotations by Einstein, as we discuss in section~\ref{sec:ontology}.  In section~\ref{sec:relativity} we show that general invariance, unlike Lorentz invariance, is not associated with a physical relativity principle.  Section~\ref{sec:discussion} contains our conclusions.  In the appendix we review the issue of Leibniz equivalence.  A much shorter presentation of these ideas, that contains only the main logical flow without any side distraction, can be found in reference~\cite{short}.

%-----------------------------------------------------------------------
\section{Invariance}
\label{sec:invariance}
\setcounter{equation}{0}
%-----------------------------------------------------------------------

In this section we discuss the symmetry properties of Einstein's field equations, and how they can be used to generate new solutions from a given one.  Since this topic is not always clear in the literature, we start discussing the more familiar, and uncontroversial, case of Lorentz symmetry.  The focus here is on {\em invariance\/} (a mathematical property of a set of differential equations~\cite{ibragimov, books}) and not on {\em relativity\/} (a physical notion).  The nontrivial connection between these two concepts will be spelled out in section~\ref{sec:relativity}.  We deliberately keep the treatment at a very elementary level; for a more sophisticated discussion, see references~\cite{ibragimov, books}.

%--------------------------------
\subsection{Lorentz invariance}
\label{subsec:lorentz}
%--------------------------------

Consider the wave equation 
\begin{equation}
\frac{\partial^2\phi}{\partial (x^1)^2}
+\frac{\partial^2\phi}{\partial (x^2)^2}
+\frac{\partial^2\phi}{\partial (x^3)^2}
-\frac{\partial^2\phi}{\partial (x^4)^2}=0\;.
\label{wave}
\end{equation}
Next, consider the coordinate transformation\footnote{Greek indices $\mu,\nu,\ldots$ run from 1 to 4.} $x^\mu\to x'^\mu$, such that $x^\mu={\Lambda^\mu}_\nu\, x'^\nu$ (or more compactly $x=\Lambda x'$), where
\begin{equation}
\left.\begin{array}{l}
{\Lambda^i}_j={\delta^i}_j+\left(\gamma-1\right)v^i\,v^j/\mbox{\boldmath
$v$}^2\\
{\Lambda^4}_i={\Lambda^i}_4=-\gamma\,v^i\\
{\Lambda^4}_4=\gamma:=\left(1-\mbox{\boldmath $v$}^2\right)^{-1/2}
\end{array}\right\}\;,
\label{Lambdaij}
\end{equation}
and $v^i$ are three real parameters such that $\mbox{\boldmath $v$}^2:=(v^1)^2+(v^2)^2+(v^3)^2<1$. It is a straightforward exercise to check that, on rewriting equation~(\ref{wave}) in terms of the new coordinates $x'^\mu$ for the function $\phi'$ defined so that $\phi'(x')=\phi(x)=\phi(\Lambda x')$, one obtains
\begin{equation}
\frac{\partial^2\phi'}{\partial (x'^1)^2}
+\frac{\partial^2\phi'}{\partial (x'^2)^2}
+\frac{\partial^2\phi'}{\partial (x'^3)^2}
-\frac{\partial^2\phi'}{\partial (x'^4)^2}=0\;.
\label{wave'}
\end{equation}
Apart from the choice of symbols, $x$ and $\phi$ in~(\ref{wave}), $x'$ and $\phi'$ in~(\ref{wave'}), the two equations are exactly the same.  As it is well known, this expresses the property of the wave equation of being invariant under Lorentz transformations of coordinates.\footnote{Of course, the wave equation is invariant under a larger set that includes also translations and rotations. We have restricted ourselves to Lorentz transformations because they are sufficient for our illustrative purposes.}

Generalising, given a set of equations in a specific coordinate system, one can find their symmetry group by looking for other coordinates in which the equations, written down explicitly, look exactly the same (see also~\cite{ibragimov, books} for the mathematical development of this idea of invariance, and~\cite{fp} for a discussion within the context of Lagrangian dynamics).  It is then possible to use this symmetry property in order to generate new solutions in a given coordinate system. For example, let $\phi(x)$ be a solution of the wave equation~(\ref{wave}). Since equation~(\ref{wave'}) is obtained by~(\ref{wave}) under the coordinate transformation $x^\mu\to x'^\mu$, it follows that $\phi'(x')=\phi(\Lambda x')$ must be a solution of~(\ref{wave'}). However, $x'$ is a mere symbol in equation~(\ref{wave'}) and in the solution $\phi'(x')$. Then, by replacing the symbol $x'^\mu$ by the symbol $x^\mu$ in~(\ref{wave'}) and in its solution, and noting that, by Lorentz invariance,~(\ref{wave'}) looks exactly the same as~(\ref{wave}), it turns out that $\phi(\Lambda x)$ is a solution of~(\ref{wave}). That is, if $\phi(x)$ is a solution of equation~(\ref{wave}), also $\bar{\phi}(x):=\phi(\Lambda x)$ is, for any $\Lambda$ that identifies a transformation in the symmetry group of the equation.\footnote{This way of generating new solutions of an equation involves an interesting interplay between the passive and active views (see reference~\cite{wald}, pp.~438--439).  By an active transformation, one can produce a new field, say $\bar{\phi}(x)$, out of an old one, $\phi(x)$. If the passive version of such a transformation is a symmetry of the field equation, then $\bar{\phi}$ is a solution whenever $\phi$ is.}

A coordinate transformation that does not constitute a symmetry of equation~(\ref{wave}) is, for example, the one defined by
\begin{equation}
\left.
\begin{array}{l}
x^1=x'^{1}\sin x'^{2}\cos x'^{3}\\
x^2=x'^{1}\sin x'^{2}\sin x'^{3}\\
x^3=x'^{1}\cos x'^{2}\\
x^4=x'^{4}
\end{array}
\right\}\;,
\label{polar}
\end{equation}
where $x'^{1}\in\mathbb{R}^{+}$, $x'^{2}\in[0,\pi]$, $x'^{3}\in[0,2\pi[$, and $x'^{4}\in\mathbb{R}$. This corresponds to using spherical polar coordinates in the three-dimensional Euclidean space.  The wave equation then takes the form
\begin{equation}
\frac{\partial^2\phi'}{\partial (x'^1)^2}
+\frac{2}{x'^1}\,\frac{\partial\phi'}{\partial x'^1}
+\frac{1}{(x'^1)^2}\,\frac{\partial^2\phi'}{\partial (x'^2)^2}
+\frac{1}{(x'^1)^2\,\tan x'^2}\,\frac{\partial\phi'}{\partial (x'^2)}
+\frac{1}{(x'^1)^2\,\sin^2 x'^2}\,\frac{\partial^2\phi'}{\partial
(x'^3)^2}-\frac{\partial^2\phi'}{\partial (x'^4)^2}=0\;.
\label{polar-wave}
\end{equation}
Clearly, equations~(\ref{wave}) and~(\ref{polar-wave}) look different and we must therefore conclude that the coordinate transformation~(\ref{polar}) does not identify a symmetry of the equation. We can thus claim that the wave equation is not invariant under general coordinate transformations.

%--------------------------------
\subsection{General invariance}
\label{subsec:general}
%--------------------------------

Contrarily to what happens for the wave equation, Einstein's field equations are invariant under general coordinate transformations.  Let us first consider Einstein's equations in empty space.  They are a set of partial differential equations for the ten functions $\g_{\mu\nu}(x)$.  When written down explicitly in coordinates, they take the form
\begin{eqnarray}
\lefteqn{\frac{\partial}{\partial x^\nu}\left(\g^{\lambda\kappa}
\left(\frac{\partial\g_{\kappa\lambda}}{\partial x^\mu}
+\frac{\partial\g_{\kappa\mu}}{\partial x^\lambda}
-\frac{\partial\g_{\mu\lambda}}{\partial x^\kappa}\right)\right)
-\frac{\partial}{\partial x^\lambda}\left(\g^{\lambda\kappa}
\left(\frac{\partial\g_{\kappa\nu}}{\partial x^\mu}
+\frac{\partial\g_{\kappa\mu}}{\partial x^\nu}
-\frac{\partial\g_{\mu\nu}}{\partial x^\kappa}\right)\right)}\nonumber\\
&&+\frac{1}{2}\,\g^{\rho\kappa}\g^{\lambda\sigma}
\left(\frac{\partial\g_{\kappa\lambda}}{\partial x^\mu}
+\frac{\partial\g_{\kappa\mu}}{\partial x^\lambda}
-\frac{\partial\g_{\mu\lambda}}{\partial x^\kappa}\right)
\left(\frac{\partial\g_{\sigma\rho}}{\partial x^\nu}
+\frac{\partial\g_{\sigma\nu}}{\partial x^\rho}
-\frac{\partial\g_{\nu\rho}}{\partial x^\sigma}\right)\nonumber\\
&&-\frac{1}{2}\,\g^{\rho\kappa}\g^{\lambda\sigma}
\left(\frac{\partial\g_{\kappa\nu}}{\partial x^\mu}
+\frac{\partial\g_{\kappa\mu}}{\partial x^\nu}
-\frac{\partial\g_{\mu\nu}}{\partial x^\kappa}\right)
\left(\frac{\partial\g_{\sigma\rho}}{\partial x^\lambda}
+\frac{\partial\g_{\sigma\lambda}}{\partial x^\rho}
-\frac{\partial\g_{\lambda\rho}}{\partial x^\sigma}\right)=0\;,
\label{einstein}
\end{eqnarray}
where $\g^{\mu\nu}(x)$ are the components of the inverse of the matrix formed by the $\g_{\mu\nu}(x)$.  Under the change of coordinates $x^\mu\to x'^\mu$, defined by $x^\mu=f^\mu(x')$, with the $f^\mu$ arbitrary differentiable functions, these equations become
\begin{eqnarray}
\lefteqn{\frac{\partial}{\partial x'^\nu}\left(\g'^{\lambda\kappa}
\left(\frac{\partial\g'_{\kappa\lambda}}{\partial x'^\mu}
+\frac{\partial\g'_{\kappa\mu}}{\partial x'^\lambda}
-\frac{\partial\g'_{\mu\lambda}}{\partial x'^\kappa}\right)\right)
-\frac{\partial}{\partial x'^\lambda}\left(\g'^{\lambda\kappa}
\left(\frac{\partial\g'_{\kappa\nu}}{\partial x'^\mu}
+\frac{\partial\g'_{\kappa\mu}}{\partial x'^\nu}
-\frac{\partial\g'_{\mu\nu}}{\partial x'^\kappa}\right)\right)}\nonumber\\
&&+\frac{1}{2}\,\g'^{\rho\kappa}\g'^{\lambda\sigma}
\left(\frac{\partial\g'_{\kappa\lambda}}{\partial x'^\mu}
+\frac{\partial\g'_{\kappa\mu}}{\partial x'^\lambda}
-\frac{\partial\g'_{\mu\lambda}}{\partial x'^\kappa}\right)
\left(\frac{\partial\g'_{\sigma\rho}}{\partial x'^\nu}
+\frac{\partial\g'_{\sigma\nu}}{\partial x'^\rho}
-\frac{\partial\g'_{\nu\rho}}{\partial x'^\sigma}\right)\nonumber\\
&&-\frac{1}{2}\,\g'^{\rho\kappa}\g'^{\lambda\sigma}
\left(\frac{\partial\g'_{\kappa\nu}}{\partial x'^\mu}
+\frac{\partial\g'_{\kappa\mu}}{\partial x'^\nu}
-\frac{\partial\g'_{\mu\nu}}{\partial x'^\kappa}\right)
\left(\frac{\partial\g'_{\sigma\rho}}{\partial x'^\lambda}
+\frac{\partial\g'_{\sigma\lambda}}{\partial x'^\rho}
-\frac{\partial\g'_{\lambda\rho}}{\partial x'^\sigma}\right)=0\;,
\label{einstein'}
\end{eqnarray}
where
\begin{equation}
\g'_{\mu\nu}(x')=\frac{\partial f^\rho(x')}{\partial x'^\mu}\,
\frac{\partial f^\sigma(x')}{\partial x'^\nu}\,
\g_{\rho\sigma}(f(x'))\;,
\label{gg'}
\end{equation}
and $\g'^{\mu\nu}(x')$ are the components of the corresponding inverse matrix.  Apart from the choice of symbols, $x$ and $\g_{\mu\nu}$ in~(\ref{einstein}), $x'$ and $\g'_{\mu\nu}$ in~(\ref{einstein'}), the two equations are exactly the same.  This is an exact counterpart of what happened for the wave equation~(\ref{wave}) under Lorentz coordinate transformations. Hence, we can claim that the group of general coordinate transformations is a symmetry group for Einstein's field equations~\cite{ibragimov}, and we say that the latter are {\em generally invariant\/}.  As we saw in the end of section~\ref{subsec:lorentz}, this is not the case for the wave equation.

One can use general invariance to generate solutions of Einstein's equations, as we did for Lorentz invariance.  Let $\g_{\mu\nu}(x)$ be a solution of~(\ref{einstein}).  Then~(\ref{gg'}) is a solution of~(\ref{einstein'}), for any choice of $f$.  On replacing $x'$ by $x$, and using general invariance, we have that the new functions
\begin{equation}
\bar{\g}_{\mu\nu}(x)=\frac{\partial f^\rho(x)}{\partial x^\mu}\,
\frac{\partial f^\sigma(x)}{\partial x^\nu}\,
\g_{\rho\sigma}(f(x))
\label{new-g}
\end{equation}
solve the original equation~(\ref{einstein}).  Of course, equation~(\ref{new-g}) can also be regarded as the chart representation of the action on the metric of a diffeomorphism $\varphi:\mathscr{M}\to\mathscr{M}$, so $\bar{\g}=\varphi_\ast\g$.  Hence the adoption of a terminology in which general relativity is diffeomorphism-invariant.

The above discussion can be easily generalised from pure gravity to the case in which matter sources are also present.  The relevant theory will then contain Einstein's equations
\begin{equation}
G_{\mu\nu}=\kappa\, T_{\mu\nu}
\label{matter}
\end{equation}
($G_{\mu\nu}$ is the Einstein tensor constructed out of $\g_{\mu\nu}$ and its first and second derivatives, $\kappa$ is the usual coupling constant of general relativity, and $T_{\mu\nu}$ is the stress-energy-momentum tensor), together with the equations of motion for other fields and the constitutive equations that express $T_{\mu\nu}$.  It is then not difficult to see that the entire set of all these equations is still generally invariant (or, equivalently, diffeomorphism-invariant).

Of course, coordinates may not always cover the whole manifold. However, this poses no problem for the previous definition of invariance. The field equations can be restricted to a topologically trivial open set of $\mathscr{M}$, and invariance can be checked for all such open regions.

%--------------------------------
\subsection{Comments}
\label{subsec:invcomm}
%--------------------------------

We have seen how to make mathematically precise the notion of invariance under transformations of the coordinates, so that it is straightforward to check whether a set of equations in the independent variables $x^\mu$ is, or is not, invariant under some coordinate transformations.  We now address various issues that have appeared in the literature.  This material is somewhat outside of the logical flow of the article, so the hurried reader may go directly to section~\ref{sec:hole} and (perhaps) come back to this section later.

%--------------------------------
\subsubsection{Invariance of field equations {\em versus\/} symmetries of solutions}
\label{subsubsec:invvssymm}
%--------------------------------

Invariance of field equations, as defined above, is sometimes erroneously conflated with invariance of solutions, \ie, symmetries of the latter.  Of course, these two notions are mathematically very distinct from each other.  For example, given any specific solution $\g_{\mu\nu}$ of Einstein's field equations, it is very easy to find functions $f^\mu$ such that $\bar{\g}_{\mu\nu}\neq\g_{\mu\nu}$.  However, it is incorrect to claim that this represents a breaking of general invariance, because such a notion concerns symmetry properties of the equations, not of their solutions.  Specific solutions, in general, do not possess the symmetries of the equations they satisfy.  For example, only trivial solutions of the wave equation~(\ref{wave}) are invariant under Lorentz transformations, but this does not affect the Lorentz invariance of~(\ref{wave}).  Of course, it is interesting in itself to study the symmetries of particular solutions in general relativity, but in this article we shall be concerned with invariance properties of field equations.

%--------------------------------
\subsubsection{General covariance {\em versus\/} general invariance}
\label{subsubsec:gcvsgi}
%--------------------------------

The mathematical property of general invariance of a system of differential equations that we discussed in section~\ref{subsec:general} is highly nontrivial, but it is not often emphasised, although many textbooks report about the ``invariance of general relativity under arbitrary diffeomorphisms''.  In fact, it is often confused with the mathematical possibility of formulating a theory using tensors --- a property with little physical relevance~\cite{kretschmann} that we shall denote {\em general covariance\/}.\footnote{Failing to appreciate the difference between invariance (a symmetry property of a set of equations) and covariance (a property of the formal apparatus used in a physical theory) has produced a huge literature. See reference~\cite{norton} for a thorough review.} For example, Einstein's equations are generally invariant, while the wave equation~(\ref{wave}) is not; however, both theories admit a tensorial formulation, so they can be written in a generally covariant way.

In order to avoid confusion about this delicate issue, we want to elaborate on it a little.  Let us define the Minkowski metric tensor specifying its components in a coordinate system in which equation~(\ref{wave}) holds through the matrix $\mbox{diag}\left(1,1,1,-1\right)$, and requiring that, under a transformation to general coordinates, the new components are given according to equation~(\ref{gg'}).  We shall denote by $\eta_{\mu\nu}$ the components of such tensor in a generic coordinate system $x^\mu$.  Furthermore, let ${\nabla\!}_\mu$ be the torsion-free covariant derivative such that ${\nabla\!}_\mu\eta_{\nu\rho}\equiv 0$.  Then the wave equation~(\ref{wave}) can be rewritten in the generally {\em covariant\/} way
\begin{equation}
\eta^{\mu\nu}{\nabla\!}_\mu{\nabla\!}_\nu\,\phi=0\;.
\label{cov-wave}
\end{equation}
However, by no means has this cosmetics enlarged the {\em invariance\/} group of the equation.  With any choice of non-Lorentzian coordinates, equation~(\ref{cov-wave}) will look, when written down explicitly, different from equation~(\ref{wave}), as we pointed out at the end of section~\ref{subsec:lorentz}.  Clearly, equations~(\ref{wave}) and~(\ref{polar-wave}) look different because the coefficients $\eta_{\mu\nu}$ are not the same in all coordinate systems.  In the tensor formulation~(\ref{cov-wave}), however, this difference is hidden.

Einstein's equations in vacuum can also be written down, of course, in the tensor form $R_{\mu\nu}=0$.  However, this fact by itself is not enough in order to guarantee general invariance, just as being able to write the wave equation in the form~(\ref{cov-wave}) does not automatically imply an invariance group larger than the Lorentz one, as we have seen. Hence, general invariance is not a fictitious property of Einstein's equations, due to the fact that they are usually written using tensors.  On the contrary, it is a precise symmetry that shows up most clearly when the equations are written down explicitly in coordinates.

%--------------------------------
\subsubsection{Parametrised field theories}
\label{subsubsec:discussion}
%--------------------------------

It is possible to turn {\em any\/} field theory into an empirically equivalent and generally invariant theory by reformulating it on a suitable manifold of parameters, which amounts to parametrising the coordinates themselves.  This trick is frequently used in analytical mechanics when time is treated as a dynamical variable by reformulating the theory on extended configuration space~\cite{Lanczos}.  We now illustrate the procedure for the wave equation~(\ref{wave}). For a general treatment, see references~\cite{kiefer, varadarajan}.

In equation~(\ref{wave}), the single unknown is the function $\phi$ of the independent variables $x^\mu$.  The dynamics of the field $\phi$ is obtained by varying a Lagrangian ${\cal L}(\phi,\partial\phi/\partial x^\mu)$.  We now construct a generally invariant theory from~(\ref{wave}), considering the coordinates $x^\mu$ as four scalar fields on some manifold $\mathscr{M}$ diffeomorphic to $\mathbb{R}^4$ on which one can introduce coordinates $\xi^\mu$, so we can write $x^\mu(\xi)$.  In order to ensure empirical equivalence with the original theory we kinematically restrict the four scalar fields $x^\mu$ to have non-vanishing Jacobian $J:=\det\left(\partial x^\mu/\partial \xi^\nu\right)$ everywhere on $\mathscr{M}$.  The ``coordinate fields'' $x^\mu:\mathscr{M}\to\mathbb{R}$ then define a diffeomorphism $\varphi:\mathscr{M}\to\mathbb{R}^4$ and we can pull-back $\phi$ on $\mathscr{M}$, obtaining the function $\bar{\phi}=\phi\circ\varphi$.  Since the Lagrangian is a scalar density of weight $+1$, the Lagrangian $\bar{\cal L}$ for the dynamical fields $\bar{\phi}$ and $x^\mu$ is:
\begin{eqnarray}
\bar{{\cal L}}\left(\bar{\phi},x^\mu,\frac{\partial\bar{\phi}}{\partial\xi^\mu},
\frac{\partial x^\mu}{\partial\xi^\nu}\right)=J{\cal L}\left(\bar{\phi},
\frac{\partial\bar{\phi}}{\partial\xi^\nu}\frac{\partial\xi^\nu}{\partial
x^\mu}\right)\;.
\label{parKG}
\end{eqnarray}
We have thus a theory of five dynamical scalar fields $\bar{\phi}(\xi),x^1(\xi),\ldots,x^4(\xi)$, and one can readily verify that the corresponding set of field equations obtained by varying $\bar{{\cal L}}$ with respect to all these fields\footnote{Note that the new Lagrangian $\bar{{\cal L}}$ does not depend on the $x^\mu$, but only on their derivatives $\partial x^\mu/\partial\xi^\nu$.  Varying $\bar{{\cal L}}$ with respect to the coordinate fields $x^\mu$ yields equations that express the conservation laws for $\phi$.} is generally invariant according to the definition given in section~\ref{subsec:general}.  The empirical equivalence of the theories~(\ref{wave}) and~(\ref{parKG}) will be discussed in section~\ref{subsubsec:kretschmannagain}.

Since the above trick of parametrising the coordinates can be carried out for any tensorial equation, we conclude that all field theories can be made generally invariant.  Merely demanding that the field equations be generally invariant puts no restrictions whatsoever on the empirical content of a theory, but only on its mathematical presentation.  The converse, namely whether any generally invariant theory can be deparametrised, is however non-trivial.  We suggest that this is the really interesting question, and we shall discuss it in section~\ref{subsubsec:depar}.   Extracting observables from a generally invariant theory --- hence making physically unambiguous predictions --- is also not an entirely trivial task, as we shall discuss at length in section~\ref{sec:observables}.

%--------------------------------
\subsubsection{Kretschmann's and Anderson's examples}
\label{subsubsec:objection}
%--------------------------------

That general invariance cannot be used as a selective principle, ruling out certain physical theories, should be obvious from the counterexample of parametrised field theories.  We now present another generally invariant reformulation of the wave equation~\cite{kretschmann, sorkin} which has been used in the literature to support similar claims.  However, we warn the reader that, for reasons that will become clear only in section~\ref{subsubsec:kretschmannagain}, this is not an appropriate counterexample.

The wave equation~(\ref{cov-wave}) in Minkowski spacetime is not generally invariant, but one can nevertheless produce a mathematically distinct theory that is apparently physically equivalent to it and which {\em is\/} generally invariant.  For this purpose, it is sufficient to introduce new unknown functions, the components of a symmetric tensor $\g_{\mu\nu}$, required to be non-degenerate with Lorentzian signature, and consider the set of equations
\begin{equation}
\left.\begin{array}{l}
\g^{\mu\nu}{\nabla\!}_\mu{\nabla\!}_\nu\phi=0\\
{R_{\mu\nu\rho}}^\sigma=0
\end{array}\right\}\,,
\label{kretschmannwave}
\end{equation}
where ${R_{\mu\nu\rho}}^\sigma$ is the Riemann tensor constructed from $\g_{\mu\nu}$.  These conditions imply that $\g_{\mu\nu}$ is a flat metric, so this theory describes a massless scalar field in Minkowski spacetime, just as equation~(\ref{wave}) does.\footnote{In fact, it does so in arbitrary coordinates, so it leads to the generally covariant version~(\ref{cov-wave}) of equation~(\ref{wave}).}  However, the system~(\ref{kretschmannwave}) is generally invariant according to the definition given above, whereas equation~(\ref{wave}) is not.  Of course, the set of equations~(\ref{kretschmannwave}) is mathematically distinct from equation~(\ref{wave}), so it is not surprising that they have different invariance properties.  But if equations~(\ref{wave}) and~(\ref{kretschmannwave}) have the same empirical content, then the principle of general invariance cannot have physical significance.  In such a case, the symmetry requirement only restricts the mathematical formulation of our physical theory.

The basic flaw of this argument is that a key ingredient of it, the empirical equivalence of the theories based on equations~(\ref{wave}) and~(\ref{kretschmannwave}), cannot be taken for granted.  Indeed, such theories are {\em not\/} empirically equivalent, unless one decides to adopt a rather unorthodox interpretation of the coordinates in equation~(\ref{wave}).  We shall return to this issue in section~\ref{subsubsec:kretschmannagain}.

In the debate revolving around this reformulation of the wave equation, Anderson has tried to defend the role of general invariance showing that, by a similar procedure, one can make the diffusion equation Lorentz-invariant~\cite{anderson-book}.  It is sufficient to replace
\begin{equation}
\frac{\partial u}{\partial x^4}=\alpha\left(\frac{\partial^2
u}{\partial (x^1)^2} +\frac{\partial^2 u}{\partial (x^2)^2}
+\frac{\partial^2 u}{\partial (x^3)^2}\right)
\label{heat}
\end{equation}
($u$ is the unknown field, and $\alpha$ is a positive coefficient) by the system
\begin{equation}
\left.\begin{array}{l}
{\displaystyle n^\mu\,\frac{\partial u}{\partial x^\mu}=
\alpha\,h^{\mu\nu}\,\frac{\partial^2 u}{\partial x^\mu\partial x^\nu}}\\
\\
\partial n^\mu/\partial x^\nu=0
\end{array}\right\}\,,
\label{syst-heat}
\end{equation}
where the $x^\mu$ are Lorentzian coordinates in Minkowski spacetime, $h^{\mu\nu}:=\eta^{\mu\nu}+n^\mu\,n^\nu$, and one requires $n^\mu$ to be a timelike unit vector.  Then, in the particular coordinate system where $n^\mu=\delta^\mu_4$, the system~(\ref{syst-heat}) reduces to equation~(\ref{heat}).  But the former is Lorentz-invariant, while the latter is not.  Hence, if one believes that general invariance is an ill-posed concept because a non-generally invariant theory can be rewritten in a generally invariant way, then the same attitude should be taken with respect to Lorentz invariance --- a very implausible conclusion.  However, on closer analysis this argument also turns out to be flawed, because the empirical equivalence between equation~(\ref{heat}) and the system~(\ref{syst-heat}) cannot be taken for granted, as we shall see in section~\ref{subsubsec:kretschmannagain}.

%--------------------------------
\subsubsection{Absolute objects}
\label{subsubsec:absolute}
%--------------------------------

Until now we have always used coordinates.  However, generally covariant equations can also be reformulated on a suitable manifold in a chart-independent --- \ie, coordinate-free --- way (see also reference~\cite{giulini}).\footnote{The possibility of writing down equations in a coordinate-free language can be taken as an alternative definition for ``general covariance'', equivalent to the one given above.}

Such a reformulation, although very appealing from a mathematical point of view, makes the notion of general invariance less transparent.  When an equation is written using a chart-independent language, it just does not make sense to analyse how its form changes under a coordinate transformation, because coordinates are not involved from the outset.  Therefore, one cannot carry out the above analysis in order to identify possible symmetries of the equations.  However, even in the coordinate-free description the symmetries of an equation are still present, and can be related to the symmetries of the so-called non-dynamical, or absolute, objects that one must introduce in order to be able to write the equation in tensorial form.

For concreteness, consider again the wave equation~(\ref{cov-wave}) in Minkowski spacetime.  It is not difficult to realise, by inspection, that the feature for which this equation fails to be generally invariant, is the fact that something in it is specified {\em a priori\/}.  More precisely, the metric coefficients in the wave equation are preassigned functions of the coordinates.  This is not the case for Einstein's equations~(\ref{einstein}), where the $\g_{\mu\nu}$ are, in fact, the unknown functions one is looking for.  In other words, the wave equation possesses some ``absolute structure'' that is not present in Einstein's theory.  The invariance group of an equation seems then to be related to the invariance group of its ``absolute objects'' (\eg, the Minkowski metric, for the wave equation in flat spacetime).  Indeed, the non-dynamical object $\eta^{\mu\nu}$ satisfies the Killing equation $\pounds_\xi\eta^{\mu\nu}=0$, where the vector $\xi$ can generate spacetime translations, spatial rotations, or boosts.  Thus, the invariance of field equations under some coordinate transformations appears to be associated with the symmetries of appropriate non-dynamical geometrical objects.\footnote{From this point of view, it is remarkable that Einstein's field equations are invariant only under differentiable coordinate transformations, so they {\em change\/} their form if non-differentiable transformations are considered.  This fact seems to suggest that general relativity does contain absolute structure after all, to be identified with differentiability and dimensionality.}

These remarks have been the basis of an attempt at defining symmetries of a theory in a chart-independent way~\cite{anderson-book}, but this turned out to be a rather involved issue, because the notion of absolute objects appears somewhat problematic~\cite{friedman, norton, anderson-book, giulini}.  Therefore, we have chosen not to make use of the coordinate-free description in our analysis.  In our view, Lorentz invariance and general invariance are most clearly understood as invariance properties of field equations under coordinate transformations.

%-----------------------------------------------------------------------
\section{The hole argument}
\label{sec:hole}
\setcounter{equation}{0}
%-----------------------------------------------------------------------

We now review the ``hole problem'' that Einstein faced when contemplating generally invariant theories in 1914~\cite{friedman, earman, norton, stachel, earman-norton, hole}.  His first reaction was to reject them as unphysical, since they apparently turn out to be physically under-deterministic.  However, the proper understanding of the problem, and its resolution, consists in a correct interpretation of the coordinates, and of the points of the manifold $\mathscr{M}$.  Although we take general relativity as an illustrative and important example, everything we say can easily be adapted to any (non-trivial) generally invariant theory.

%--------------------------------
\subsection{Interpretations of coordinates and points of the manifold}
\label{subsec:coordinates}
%--------------------------------

Following Einstein's 1905 article, the variables $x^1,\ldots,x^4$ used in special relativity are commonly interpreted as operationally well-defined quantities, that refer to readings on rulers and clocks, or to the arrival of some light signal.  Thus, these variables represent the values of physical quantities that one uses in order to give a clear operational meaning to the notions of position and time.  Consider, for example, the scalar field $\phi$ in equation~(\ref{wave}).  In the expression $\phi(x^1,x^2,x^3,x^4)$, the quantities $x^1,x^2,x^3$ can be taken to express actual readings along three orthogonal rigid rulers, and $x^4$ an actual reading on a clock.  Hence, $\phi(x^1,x^2,x^3,x^4)$ does not represent the value of the scalar field at some abstract ``instant of time'' and ``point of space'', but rather the value of the field in correspondence with some other physical quantities. Of course, such an interpretation can also be adopted in Newtonian mechanics, although one seldom states this explicitly.

There are many ways to introduce coordinates that have an operational meaning, although the construction of accurate clocks and rulers is a non-trivial task, both practically and theoretically.  In what follows we shall not care how this is done, but simply assume that {\em some\/} operational definition is adopted.  We shall refer to the material system used for such a purpose as the {\em coordinate scaffolding\/}, without further specification about its constitution.  (In particular, a coordinate scaffolding is not necessarily some artificial object built by an experimentalist; one could equally well use a configuration that is already present in Nature.)  In general, an event is characterised by the values of a large number of physical quantities of the scaffolding, most of which are, however, irrelevant for the purpose of extracting coordinates.  The coordinates will thus be obtained by generating four real numbers out of these physical quantities, so $x^1,\ldots,x^4$ can be viewed, more formally, as real functions on the set of events.  If one now uses these functions to define a chart on a manifold, the latter is easily interpreted as the set of all events, identified through convenient physical readings on the scaffolding.

This operational interpretation of the coordinates is not the only possible one. One can adopt an alternative interpretation of the variables $x^\mu$ as mere mathematical parameters, devoid of any operational significance.  In the above example of the scalar field, this means that a functional relation between the parameter values $x^\mu$ and the values of $\phi$ cannot be experimentally established.  The parameters $x^\mu$ then do not identify an operationally well-defined position in space and time, although they can be regarded as defining a chart on an abstract manifold.  Such a manifold should not, however, be confused with the space of all events, which requires the presence of physical fields for its very definition.\footnote{This important conceptual distinction is normally not made in the literature.}

It is remarkable that theories like Newtonian mechanics and special relativity allow for two alternative interpretations of the coordinates $x^\mu$,  so that these theories admit both a concrete, operational formulation, and an abstract one.  Similarly, there are two possible interpretations for the spacetime manifold that is introduced when these theories are formulated in a coordinate-free way.  One can imagine it to be the set of all events, so that it is somehow endowed with physical properties that allow one to give an operational meaning to its points.  Or one could take it to be an abstract space of parameters, with no operational meaning whatsoever.  We now show that general invariance forces one to adopt the second interpretation --- so, in particular, in general relativity coordinates and manifold points do not have an operational meaning.

%--------------------------------
\subsection{The $x^\mu$ are not readings}
\label{subsec:hole-coord}
%--------------------------------

Consider a solution $\g_{\mu\nu}(x)$ of Einstein's equations~(\ref{einstein}) in some coordinates $x^\mu$.  Within a suitable open region of the coordinate domain, $\g_{\mu\nu}(x)$ can be regarded as the solution of an initial value problem formulated on a three-dimensional hypersurface.  We can now use the general invariance of Einstein's equations to generate, starting from $\g_{\mu\nu}(x)$, a different solution $\bar{\g}_{\mu\nu}(x)$ that satisfies the same initial value problem.  For this purpose, it is sufficient to choose the functions $f^\mu$ in equation~(\ref{new-g}) such that they coincide with the identity only in a neighbourhood of the initial hypersurface, while they differ from it elsewhere.

This simple observation has far-reaching consequences.  We shall now adopt the interpretation in which the values of the $x^\mu$ represent physical readings.  We will then reach an untenable conclusion, so the operational interpretation of the coordinates must be dropped, by a {\em reductio ad absurdum\/}.

By assumption, a specific value of the $x^\mu$ corresponds to operationally well-defined readings, so one expects any measurable quantity, for example the Kretschmann scalar $R_{\mu\nu\rho\sigma}R^{\mu\nu\rho\sigma}$, to be expressed by a unique well-defined scalar function of the $x^\mu$, say $s(x)$.\footnote{For concreteness, and without loss of generality, we shall often refer to the Kretschmann scalar, but the argument obviously applies to any non-constant scalar function constructed from the metric.  Of course, theories in which one can construct only constant scalars, even if generally invariant escape the argument in the form given below.  (This is the case, for example, of a spacetime theory with field equations ${R_{\mu\nu\rho}}^\sigma=0$.)  However, such theories are physically uninteresting.  Moreover, if the coordinates have an operational meaning, even the {\em components\/} of tensors, such as $\g_{\mu\nu}$, are empirically accessible quantities, so one could run a hole argument using them instead of scalars.}  Indeed, in any actual experiment only one correspondence between the values of the Kretsch\-mann scalar and those of the (by assumption) operationally well-defined readings $x^\mu$ will be found.  However, as stated above, $\g_{\mu\nu}(x)$ and $\bar{\g}_{\mu\nu}(x)$ are both solutions of the same initial value problem, and in these two mathematically distinct solutions the functional dependence of the Kretschmann scalar on the coordinates are $s(x)$ and $\bar{s}(x):=s(f(x))$, respectively.  Since $s(x)\neq s(f(x))$ in general and so $s(x)\neq\bar{s}(x)$, it follows that, because of the general invariance of Einstein's equations, general relativity does not predict a unique value of the Kretschmann scalar for given values of the $x^\mu$.  We have thus reached the conclusion that general relativity is unable to make unique {\em empirical\/} predictions.  In other words, thinking of the $x^\mu$ as actual readings, Einstein's equations imply physical under-determinism.

Failing to appreciate the crucial distinction between the parameters $x^\mu$ employed in general relativity, and the physical coordinates commonly used in special relativity, has caused much confusion.  Einstein himself made such a mistake when he rejected generally invariant theories, believing that they had to be under-deterministic because of the hole argument~\cite{norton}.  However, another way out is to drop the assumption that the coordinates $x^\mu$ have any operational meaning --- that is, that they correspond to readings of some sort.  They are just {\em mathematical parameters\/}.  Note that this should not be interpreted in the trivial sense that charts on a manifold are arbitrary, because, given a chart, there is a one-to-one correspondence between a manifold point and the coordinates.  Hence, what we are saying is actually that the {\em manifold points\/} lack operational significance.  In this respect, they are just like their coordinates.  This claim can be strengthened by reformulating the hole argument in a coordinate-free language.

%--------------------------------
\subsection{Points of $\mathscr{M}$ are not events}
\label{subsec:hole-abstr}
%--------------------------------

It is tempting to respond to the above argument by pointing out that the use of coordinates is optional, and that one could adopt a coordinate-free description on some manifold $\mathscr{M}$.  It might then be thought that it is the points of this manifold that identify operationally well-defined positions in space and instants of time, rather than four arbitrarily introduced scalar fields on $\mathscr{M}$ --- the coordinates.  Indeed, it is commonplace in textbooks to associate points of $\mathscr{M}$ with physical events (see, \eg, reference~\cite{mtw}).  We shall now see that in this coordinate-free description, the manifold points cannot correspond to operationally well-defined events.  The argument runs similarly to the one presented in section~\ref{subsec:hole-coord}.

Let a spacetime model $(\mathscr{M},\g)$ be a solution of Einstein's equations in vacuum.  Consider a diffeomorphism $\varphi:\mathscr{M}\to \mathscr{M}$ that coincides with the identity in a neighbourhood of a spacelike hypersurface $\mathscr{S}$, and which differs from the identity somewhere in $D^+(\mathscr{S})$, the future domain of dependence of $\mathscr{S}$ defined according to $\g$.  Let $\overline{D}^+(\mathscr{S})$ be the future domain of dependence of $\mathscr{S}$ according to $\bar{\g}=\varphi_\ast\g$, and define $\mathscr{D}:=D^+(\mathscr{S})\cap \overline{D}^+(\mathscr{S})$.  Then, by general invariance, both $(\mathscr{D},\g)$ and $(\mathscr{D},\bar{\g})$ are solutions of the same initial value problem for Einstein's equations.  However, by construction of $\varphi$ there are points $p\in \mathscr{D}$ for which $\g(p)\neq\bar{\g}(p)$.  At these points, there are then observable quantities for which the theory does not make a unique prediction. Indeed, for a generic scalar (for example, the Kretschmann scalar) we have the possible values $s(p)$ and $\bar{s}(p)=\varphi_{\ast} s(p)=s(\varphi^{-1}(p))$.  In general, $s(p)\neq \varphi_{\ast}s(p)$, so general invariance does not allow for a well-defined correspondence between points of $\mathscr{M}$ and values of $s$.  On the other hand, there is such a correspondence between physical events and values of scalar quantities.  Therefore, points of $\mathscr{M}$ cannot represent events.  The manifold $\mathscr{M}$ must be thought of as a purely abstract entity, whose points possess no physical quality that could allow one to identify them, \ie, they necessarily represent empirically inaccessible structure in general relativity.

Hence, in a theory whose equations are generally invariant the assumption that points of the manifold have operational meaning (or that coordinates can be identified with readings) becomes untenable, unless one is ready to accept that the theory does not make unique empirical predictions (under-determinism).  In such a theory one {\em must\/} adopt the point of view in which neither the coordinates nor the manifold points have an operational interpretation.

Interestingly, having reached this conclusion one can adapt a well-known argument by Leibniz (see Appendix) and argue that in general relativity (actually, in any generally invariant theory) a physical situation is identified by a whole equivalence class of spacetime models, all diffeomorphic to each other.\footnote{Not surprisingly, this point is well-known and well understood, at the technical level, within the context of the general relativistic Cauchy problem.  See, {\em e.g.\/}, reference~\cite{he}, pp.~227--228.}  Although the physical equivalence between diffeomorphic models is a general feature of physical theories, in some situations (\eg, Newtonian mechanics, special relativity) one can adopt the ``readings interpretation'', thus making Leibniz's reasoning inapplicable from the outset.  Indeed, if the manifold points are assumed to designate operationally well-defined events, then two models related by a non-trivial diffeomorphism describe two physically distinct situations.\footnote{In particular, if points of $\mathscr{M}$ are physically identifiable, the two metrics $\g$ and $\varphi_\ast\g$ can correspond to arbitrarily different distances on $\mathscr{M}$, so they do not describe the same geometry.}  However, the general invariance of Einstein's equations makes Leibniz's argument compelling in general relativity.  These observations are important, because they clarify the subtle connection between invariance properties of equations and physical relativity principles, that will be discussed in detail in section~\ref{sec:relativity}.  

Rejecting the readings interpretation raises a number of foundational questions. If events cannot be identified with points of the manifold $\mathscr{M}$, how are they represented in a generally invariant spacetime theory? And if points of the manifold do not correspond to something empirically accessible, how does one extract observable quantities from the theory? As we shall see in the next two sections, these problems are deeply connected.

%-----------------------------------------------------------------------
\section{Observables}
\label{sec:observables}
\setcounter{equation}{0}
%-----------------------------------------------------------------------

It is commonly accepted that, in a given physical situation, a quantity is described by a specific field on the manifold $\mathscr{M}$ (or by its representative in a chart, \ie, a function of the coordinates $x^\mu$).  However, in a generally invariant theory the manifold is unobservable (and the $x^\mu$ cannot be thought of as readings of physical objects). It is then clear that, in any model $(\mathscr{M},T)$, statements referring to the value of a given field at some point of $\mathscr{M}$ are, if taken by themselves, physically empty.  How can one then extract from $(\mathscr{M},T)$ its observable properties?  What is the physical content of such a model, if the $x^\mu$ are just parameters without an intrinsic physical meaning, labeling points of $\mathscr{M}$ which also have no intrinsic physical meaning?

%--------------------------------
\subsection{The problem}
\label{subsec:probobs}
%--------------------------------

By general invariance, for each initial value problem there is an entire class of mathematically distinct solutions.  If general relativity is capable of making clear-cut predictions, all these solutions must represent the same physical situation.

This is not different from what happens in Maxwell's theory when the electromagnetic field is represented by a vector potential $A_\mu$.  Even in such a case there is an entire class of vector potentials that satisfy the same initial condition, and that must be taken to represent the same physical system --- hence, they are physically equivalent.  An observable is a quantity that can be extracted from a specific solution, and that is the same no matter which member of the equivalence class one is considering. Since any two equivalent vector potentials $A_\mu$ and $\bar{A}_\mu$ differ by the gradient of a function $\theta$, that is $\bar{A}_\mu =A_\mu +\partial_\mu\theta$, the quantity defined as $F_{\mu\nu}[A] :=\partial_\mu A_\nu -\partial_\nu A_\mu$ turns out to be an observable, because $F_{\mu\nu}[A]=F_{\mu\nu}[\bar{A}]$.

In general relativity, two members $\g$ and $\bar{\g}$ of the equivalence class are related by an arbitrary diffeomorphism, $\bar{\g}=\varphi_\ast\g$, so any corresponding observable $\cal O$ must be such that ${\cal O}[\g]={\cal O}[\varphi_\ast\g]$. If ${\cal O}[\g]$ is assumed to be a scalar $s$ defined on the manifold $\mathscr{M}$, one has that $s[\varphi_\ast\g]=\varphi_\ast s[\g]$, so one arrives at the conclusion that $s[\g]=\varphi_\ast s[\g]$, for any diffeomorphism $\varphi$. The only scalars that satisfy this condition are constants.  We must therefore conclude that observables in general relativity cannot be represented by non-trivial scalar functions on $\mathscr{M}$. This conclusion generalizes to tensor fields $T$ on $\mathscr{M}$, because the condition $T[\g]=T[\varphi_{\ast}\g]=\varphi_{\ast}T[\g]$ is satisfied only by trivial combinations of Kronecker deltas and constants. This may seem awkward, so let us elaborate on this subtle issue.

As we saw in section~\ref{subsec:coordinates}, special relativity admits an interpretation in which the coordinates $x^\mu$ are operationally definable, as actual readings on a coordinate scaffolding.  For example, when we measure a scalar field $\phi(x)$ in special relativity, we do not just record the different values that it takes (the image of $\phi$). Rather, we record the value of $\phi$ at a specific spacetime event, operationally defined by the values of the $x^\mu$ on the coordinate scaffolding.  This measurement is then repeated for many events (in principle, for all those belonging to some open region $\mathscr{V}\subseteq\mathbb{R}^4$).  In this way not only is the set of field values recorded, but also the functional relation between the values of $\phi$ and the spacetime events defined by $x^\mu$.  That is, we measure the {\em function\/} $\phi:\mathscr{V}\to\mathbb{R}$, not merely its image.

In contrast, general relativity does not admit such an operational interpretation of the coordinates $x^\mu$.  Thus, a functional relation between the parameter values $x^\mu$ and the values of $\phi$ cannot be experimentally established; that is, observables in general relativity cannot be thought of as functions on $\mathscr{M}$.  The following construction of local observables circumvents this difficulty.

%--------------------------------
\subsection{Constructing local observables}
\label{subsec:localobservables}
%--------------------------------

We have seen that observables cannot be thought of as non-trivial functions (scalar or tensorial) on $\mathscr{M}$. In coordinate language this means that they cannot be coordinate-dependent. Thus, in order to construct observables from a given spacetime model we must, in one way or another, eliminate the coordinate dependence.  An obvious way to do this is by integrating a scalar density over the whole manifold $\mathscr{M}$. However, this will hardly correspond to any local observable. In what follows we outline a particular way to construct local observables in general relativity (see also references~\cite{bergmann, dewitt}).  First we show how to construct them from scalar functions, then we generalise the procedure to arbitrary tensors.

Assume that we are given a specific spacetime model\footnote{Here we allow for possible non-gravitational fields, but the discussion is completely general, and applies to the case of pure gravity as well.} $(\mathscr{M},T)$.  Although the correspondence between manifold points (or the parameters $x^\mu$) and values of physical fields is not observable, the correspondence between values of physical fields and values of other physical fields --- point-coincidences --- {\em is\/} physically meaningful.  Indeed, this correspondence contains {\em everything\/} one needs to know, as it was already clearly expressed by Einstein~\cite{einstein-1916}:
\begin{quote}
{\em All our space-time verifications invariably amount to a determination of space-time coincidences.  [...]  The introduction of a system of reference serves no other purpose than to facilitate the description of the totality of such coincidences.\/}
\end{quote}
To define observable quantities, we then need just to construct, among all the physical and geometrical fields in our model, four {\em coordinate fields\/}, and express any other quantity in terms of these.   More precisely, suppose we can construct four scalar fields $q^\mu$ from the given model.  We stress that these fields are not introduced {\em in addition\/} to the known fields, but constructed from them; that is, each of the $q^\mu$ can be expressed entirely in terms of $T$.  Their nature is irrelevant for the discussion that follows, as long as they represent experimentally accessible quantities.  They could simply be physical fields (also test fields, or scalar quantities constructed out of the metric --- curvature invariants~\cite{bergmann}), combinations of physical fields, or even objects whose definition involves conventions like ``the number of atoms along one side of a straight iron bar, counted starting from a given end''.   Also, the $q^\mu$ need not be fundamental, but can just represent phenomenological properties of matter.  We assume that these four scalar fields are invertible,\footnote{If the $q^\mu$ were not invertible, they would not be appropriate as coordinate fields.  Note, however, that the hypothesis of invertibility can be satisfied only locally, and only {\em after\/} one has a specific model of spacetime.  There is no way to choose {\em a priori\/} four fields $q^\mu$ that can be used everywhere in a given model, and for all models.  See section~\ref{subsec:invertprob} for more details about this issue.} \ie, that in some open set $\mathscr{U}\subseteq \mathscr{M}$ the relationship between the $q^\mu$ and the parameters $x^\mu$ is such that $\det\left(\partial q^\mu/\partial x^\nu\right)\neq 0$. The image of $\mathscr{U}$ under the action of the map $q:=(q^1,q^2,q^3,q^4)$ is then a subset ${\cal Q}\subseteq\mathbb{R}^4$ with non-zero measure, and both $q:\mathscr{U}\to {\cal Q}$ and $q^{-1}:{\cal Q}\to\mathscr{U}$ are well-defined.

%--------------------------------
\subsubsection{Scalars}
\label{subsubsec:scalars}
%--------------------------------

Consider a scalar field $s:\mathscr{M}\to \mathbb{R}$ (for example, the Kretschmann scalar).  By composing $s$ and $q^{-1}$ we obtain the function $\tilde{s}=s\circ q^{-1}:{\cal Q}\to \mathbb{R}$ or, using coordinates,\footnote{With some abuse of notation, we denote by the same symbol $s$ both the function defined on $\mathscr{M}$ and its coordinate representation in a chart.  Also, we denote by $x(q)$ the values of the coordinates corresponding to the values $q$ of the physical coordinate fields.} $\tilde{s}(q)=s(x(q))$. This no longer contains associations between the measurable field values and the unobservable spacetime points, but only between measurable field values and other measurable quantities --- the $q^\mu$.  Hence, contrary to what happened for $s$ (or its coordinate representation), the function $\tilde{s}$ is observable (often called a {\em relational observable\/}, in order to stress that it only refers to field configurations, and not to some metaphysical ``spacetime'' independent of the latter).

In order to double-check this important conclusion, let us also verify that $\tilde{s}(q)$ does not depend on the member of the equivalence class of spacetime models it is constructed from.  Two such models $(\mathscr{M},T)$ and $(\mathscr{M}',T')$ are related by a diffeomorphism $\varphi:\mathscr{M}\to \mathscr{M}'$.  (Here $\mathscr{M}'$ could either be $\mathscr{M}$ itself or a different manifold with the same dimension and topology.)  In the second model the four scalar fields are given by $q'^\mu=q^\mu\circ\varphi^{-1}$. Similarly, $s'=s\circ\varphi^{-1}$. The candidate observable is $\tilde{s}'=s'\circ q'^{-1}=s\circ\varphi^{-1}\circ(q\circ\varphi^{-1})^{-1}= s\circ\varphi^{-1}\circ\varphi\circ q^{-1}=s\circ q^{-1}=\tilde{s}$.  So the quantity $\tilde{s}$ is indeed diffeomorphism-invariant, and independent of which physically equivalent model we construct it from --- it does not depend on the choice of the member in the Leibniz equivalence class (see Appendix).  In the coordinate language, had we started from the same model expressed in different coordinates, $s'(x')$, the relationship between the values of the coordinate fields and $x'$ would be given by functions $q'^\mu(x')$ that also differ from the $q^\mu(x)$, in such a way as to produce the same function $\tilde{s}(q)$.

Note that the observable $\tilde{s}$ is defined as a function on $\cal Q$ and not on $\mathscr{M}$.  This agrees with the conclusion in section~\ref{subsec:probobs}, that observable quantities in general relativity cannot be represented by non-constant functions {\em on the manifold\/} $\mathscr{M}$.

It has been recently demonstrated~\cite{dittrich} that not only are the relational observables diffeo\-morphism-invariant; they are also Dirac observables in the technical sense, {\em i.e.\/}, they weakly commute with the first-class constraints.

%--------------------------------
\subsubsection{Vectors and tensors}
\label{subsubsec:vectors}
%--------------------------------

We now address vectors and tensors.  First we note that the four scalar fields define the tetrad\footnote{Note the different meaning of the indices in $e^\mu_\nu$: Both $\mu$ and $\nu$ run from 1 to 4, but whereas $\mu$ labels the elements in a set of scalars, $\nu$ labels coordinates in a chart.  Hence, $e^\mu_\nu$ is the $\nu$-th component of the $\mu$-th element of the tetrad.  Similar remarks apply to $f_\mu^\nu$ defined below (the $\nu$-th component of the $\mu$-th element of the inverse tetrad) and to the scalars defined in equations~(\ref{onescalar})--(\ref{threescalars}).} $e^\mu_\nu:=\partial q^\mu /\partial x^\nu$.  Because of invertibility, this tetrad is non-degenerate (\ie, $\det(e^\mu_\nu)\neq 0$), so we can also introduce the inverse tetrad $f_\mu^\nu=\partial x^\nu /\partial q^\mu$ such that $e^\mu_\nu f_\rho^\nu=\delta^\mu_\rho$.

Now we can use these tetrads to construct scalars out of contravariant, covariant, or mixed tensor objects. For example:
\begin{eqnarray}
\bar{V}^\mu(x)&=&e^\mu_\nu(x)V^\nu(x)\;;\label{onescalar}\\
\bar{\g}_{\mu\nu}(x)&=&f^\rho_\mu(x) f^\sigma_\nu(x)
\g_{\rho\sigma}(x)\;;\label{twoscalars}\\
\mbox{$\bar{R}_{\mu\nu\rho}$}^\sigma(x)&=& f_\mu^\lambda(x) f_\nu^\kappa(x)
f_\rho^\tau(x) e^\sigma_\eta(x)
{R_{\lambda\kappa\tau}}^\eta(x)\;.
\label{threescalars}
\end{eqnarray}
These sets of scalars depend on the coordinates, so they are not yet observable quantities. However, observables can now easily be constructed as discussed in section~\ref{subsubsec:scalars}, by making use of the inverse relation $x=x(q)$.  In this way we obtain the set of observables:
\begin{eqnarray}
\widetilde{V}^\mu(q)&=&\bar{V}^\mu(x(q))\;;\\
\tilde{\g}_{\mu\nu}(q)&=&\bar{\g}_{\mu\nu}(x(q))\;;\\
{\widetilde{R}{_{\mu\nu\rho}}}^\sigma(q)&
=&\mbox{$\bar{R}_{\mu\nu\rho}$}^\sigma(x(q))\;.
\end{eqnarray}
These objects correspond to quantities that can be measured, at least in principle.  In fact, {\em all\/} the observable quantities of the theory --- the point-coincidences --- can be generated in this way, possibly switching to other coordinate fields whenever invertibility fails.  From them, one can read directly the values of physical quantities corresponding to the measured values of the coordinate fields.

In synthesis, although any specific spacetime model can be represented as a pair $(\mathscr{M},T)$, the functions expressing the values of fields at points of $\mathscr{M}$ have no direct physical meaning.  However, one can isolate some functions $q^\mu$ --- the coordinate fields --- and use them to express suitable combinations of the components of the various tensor fields in $T$, as functions of the $q^\mu$.\footnote{Of course, one could also regard the $q^\mu$ as just a specific choice of chart.  But contrarily to what happens for the coordinates $x^\mu$, the $q^\mu$ are constructed from the physical fields and not postulated in addition.}  These functions capture all the physically relevant (\ie, gauge-invariant) properties of the model.

The hole problem does not arise if one works only with the observable quantities constructed as described above.  Note that this simple and natural resolution is based on the {\em elimination\/} of the unphysical parameters $x^\mu$ (or of the manifold points), and should not be confused with other suggestions in which points of $\mathscr{M}$ are instead purportedly {\em individuated\/} by the physical fields~\cite{stachel}.

%-----------------------------------------------------------------------
\subsection{Remarks}
\label{subsec:obscomm}
%-----------------------------------------------------------------------

Now that we have seen how to construct the observables in a generally invariant theory, we can reconsider some issues left open from sections~\ref{sec:invariance} and~\ref{sec:hole}.

%-----------------------------------------------------------------------
\subsubsection{Predictions in general relativity}
\label{subsubsec:predictions}
%-----------------------------------------------------------------------

The problem of observables as formulated above concerns classical general relativity.  However, ever since the theory was first formulated in 1916, observable predictions have been made, {\em apparently without\/} going through the above procedure for constructing local observables.  We now clarify this potentially confusing situation.  Again, as we shall see, the issue revolves around the interpretation of coordinates.

For concreteness, consider two theories, one of which describes pure gravity, while the other describes gravity plus four massless scalar fields $q^\mu$ whose back-reaction on the gravitational field is negligible.  We take the field equations for these theories to be, respectively,
\begin{equation}
R_{\mu\nu}=0
\label{puregrav}
\end{equation}
and
\begin{equation}
\left.\begin{array}{l}
R_{\mu\nu}=0\\
\g^{\mu\nu}\nabla_{\!\mu}\nabla_{\!\nu}\,q^\rho=0
\end{array}\right\}\;.
\label{gravplus4scalars}
\end{equation}
Both theories are generally invariant according to the definition given in section~\ref{subsec:general}.

The former theory describes the gravitational field alone, with no additional matter fields.  Because of general invariance the symbols $x^\mu$ cannot be interpreted as representing the readings on a physical coordinate scaffolding.  More generally, there is nothing in this theory that can be interpreted as representing the readings on a material scaffolding --- in particular, there is no laboratory equipment (GPS devices, {\em et cetera\/}).

The latter is a theory of the gravitational field and four scalar fields.  In contrast to the former one, this theory does not exclude the presence of additional matter fields.  Indeed, the scalar fields can be understood precisely as representing the readings on some material coordinate scaffolding.  It is important to note that these fields represent test matter, since they do not back-react on the metrical field.

Imagine now that we solve equation~(\ref{puregrav}) working in coordinates that satisfy the harmonic condition
\begin{equation}
\g^{\mu\nu}\nabla_{\!\mu}\nabla_{\!\nu}\, x^\rho=0\;,
\end{equation}
thus obtaining some solution $\g^{(1)}_{\mu\nu}(x)$.  Similarly, imagine that we obtain a solution of~(\ref{gravplus4scalars}): $\g^{(2)}_{\mu\nu}(x)$ and $q^\mu(x)$, in some coordinate system $x^\mu$.  Because of general invariance, we know that in both cases the symbols $x^\mu$ cannot have operational significance.  However, assuming invertibility, in the second theory one can construct the local observables $\tilde{\g}^{(2)}_{\mu\nu}(q)$.  By carefully choosing initial data for the four scalar fields $q^\mu$, one can manage to have the same functional form for the objects $\g^{(1)}_{\mu\nu}(x)$ and $\tilde{\g}^{(2)}_{\mu\nu}(q)$.

Even though there is no functional difference, $\g^{(1)}_{\mu\nu}(x)$ is not an observable in the first theory, whereas $\tilde{\g}^{(2)}_{\mu\nu}(q)$ is an observable in the second one.  Thus, depending on which theory the metric components come from, they are, or are not, observable.

It seems reasonable that the above generalizes. Given any solution $\g^{(1)}_{\mu\nu}(x)$ to equation~(\ref{puregrav}) in some specific coordinate system, there is always a different theory involving more fields (which do not back-react on the metrical field and from which the coordinate fields $q^\mu(x)$ will be constructed) such that the observable $\tilde{\g}^{(2)}_{\mu\nu}(q)$ is functionally identical to $\g^{(1)}_{\mu\nu}(x)$.  How to build such a scaffolding in practice, depends on the specific form of the metric $\g^{(2)}_{\mu\nu}(x)$.

This makes it possible to shortcut the above construction of observables. If $\g^{(1)}_{\mu\nu}(x)$ is a solution of equation~(\ref{puregrav}) in some coordinate system $x^\mu$, just {\em imagine\/} that these components had been obtained as the set of observables $\tilde{\g}^{(2)}_{\mu\nu}(q)$ from a {\em different\/} theory with additional fields representing a coordinate scaffolding that does not back-react on the metric.  Then make use of the explicit form of the metric to figure out how to construct the appropriate material coordinate scaffolding.

Note that it is the empirical predictions of the second theory that one tests in general relativity ({\em e.g.\/}, through light bending, gravitational redshift, and time dilation).  The first theory comes with a commitment of there being no material coordinate scaffolding, and in particular no laboratory equipment.  As such, it is rather uninteresting for experimentalists!

This analysis shows how harmless the problem of observables is in {\em classical\/} general relativity.  It is still not entirely clear to what degree the issue is serious in different approaches to quantum gravity.

%--------------------------------
\subsubsection{On empirical equivalence}
\label{subsubsec:kretschmannagain}
%--------------------------------

If there is a ``natural'' one-to-one correspondence between the observables of two mathematically distinct theories, they are empirically equivalent.  For example, in the case of the wave equation~(\ref{wave}) and the parametrised field theory~(\ref{parKG}), the empirical equivalence is immediate.  The observables in the theory~(\ref{wave}) are the functions $\phi(x)$, while in the parametrised field theory~(\ref{parKG}) they are given by the composite functions $\tilde{\phi}(x)=\bar{\phi}\circ \varphi^{-1}(x)=\phi(x)$, so the one-to-one correspondence is obvious.

However, the empirical equivalence between the theories~(\ref{wave}) and~(\ref{kretschmannwave}) is non-trivial.  Since the set of equations~(\ref{kretschmannwave}) is generally invariant, the coordinates in it cannot have operational significance.  This means that the symbols $x^\mu$ cannot represent readings on some coordinate scaffolding, and the function $\phi(x)$ is not observable.  However, equation~(\ref{wave}) is not generally invariant and {\em does\/} admit an operational interpretation of the coordinates.  In this case the function $\phi(x)$ is observable.

In order to establish empirical equivalence between these theories, one would have to construct a set of (diffeomorphism-invariant) observables in the theory~(\ref{kretschmannwave}), and exhibit a one-to-one correspondence between these observables and the solutions of equation~(\ref{wave}).  But such a ``natural'' correspondence does not exist unless extra structure is added to the theory, {\em e.g.\/}, more scalar fields representing readings on a coordinate scaffolding.

Conversely, if one adopts the non-operational interpretation of the coordinates in~(\ref{wave}), then the functions $\phi(x)$ that solve it are not observable.  The empirical content, as regards the scalar field $\phi(x)$, is then presumably the same for both theories~(\ref{wave}) and~(\ref{kretschmannwave}), but these are now physically rather uninteresting, because it is impossible to extract physical predictions from them.  For example, whereas equation~(\ref{wave}) gives a perfectly viable and physically significant theory within the readings interpretation of the $x^\mu$, it is very incomplete if one adopts instead the non-operational interpretation, because the functional relationship between the $x^\mu$ and $\phi$ has no empirical counterpart.  In order to extract physical predictions, one should then supplement equation~(\ref{wave}) by other equations describing the scaffolding, then eliminate the $x^\mu$ in order to remain only with point-coincidences between readings on the scaffolding and values of $\phi$.  This is never done in practice, and an operational interpretation of the coordinates in equation~(\ref{wave}) is mostly taken for granted by physicists.

It should now be clear that equation~(\ref{heat}) and the system~(\ref{syst-heat}) actually describe different physical theories if we adopt the readings interpretation of the coordinates.  Indeed, the empirical equivalence between them holds only for a specific solution of $n^\mu$, and restricting the solution space in such a way, Lorentz invariance is broken. If the readings interpretation is dropped the theories are presumably empirically equivalent but, of course, also physically uninteresting.

%-----------------------------------------------------------------------
\subsubsection{On the deparametrisability of generally invariant theories}
\label{subsubsec:depar}
%-----------------------------------------------------------------------

The fact that one could make any theory generally invariant without changing its empirical content, simply parametrising it, shows that general invariance is not a selective principle, as we already pointed out in section~\ref{subsubsec:discussion}.  We now want to investigate the converse: Given a generally invariant theory, can it be {\em deparametrised\/}~\cite{kiefer}?  That is, can it be turned into an empirically equivalent theory that is not a gauge theory with respect to diffeomorpishms of the parameter manifold $\mathscr{M}$?  In particular, we are interested in the possibility that general relativity could be deparametrised.  This issue was addressed in reference~\cite{kuchar72}, which contains a discussion of the process of deparametrisation for general relativity under the {\em assumption\/} that such deparametrisation is possible, and, more recently, in~\cite{BrownKuchar95}, where the possibility of deparametrising general relativity with special matter content and in special situations was investigated.

Obviously, the parametrised scalar field theory of section~\ref{subsubsec:discussion} can be deparametrised.  The coordinate fields $x^\mu(\xi)$ define a diffeomorphism $\varphi$ between $\mathscr{M}$ and $\mathbb{R}^4$. Thus, we can push-out the fields $\bar{\phi}$ and $x^\mu$ on $\mathbb{R}^4$, and change the Langrangian density $\bar{\cal L}$ correspondingly.  In doing so, the coordinate fields $x^\mu$ again become a set of four non-dynamical scalar fields on $\mathbb{R}^4$, and the Lagrangian becomes the original one, ${\cal L}$.  The dynamical variable is now the single function $\phi(x)$. This function is, of course, also the observable of the parametrised theory~(\ref{parKG}), namely $\tilde{\phi} =\bar{\phi}\circ\varphi^{-1}=\phi$.

A little reflection reveals that what makes deparametrisation possible is the existence of four dynamical scalar fields which are everywhere invertible {\em for all\/} kinematically allowed models.  If this is the case, these four scalar fields define a diffeomorphism between $\mathscr{M}$ and $\mathbb{R}^4$. One can then proceed as in the above case of the parametrised scalar field theory: Push-out all dynamical fields on $\mathbb{R}^4$, and transform the Lagrangian to obtain the dynamics for the dynamical fields on $\mathbb{R}^4$.  The dynamical equations so obtained will not be generally invariant.

In the parametrised field theory of section~\ref{subsubsec:discussion} four everywhere invertible scalar fields, the coordinate fields $x^\mu(\xi)$, were available {\em by construction\/}.  However, in a theory like general relativity it is not possible to extract some four everywhere invertible scalar fields.  This can already be expected by the fact that general relativity does not contain four preferred scalar fields, so its structure differs fundamentally from the one of parametrised field theories.  Moreover, even if one could identify such fields, it is not obvious that they could not take the same values in two (or more) points of the coordinate domain.  In addition, if the topology is non-trivial ({\em i.e.\/}, $\mathscr{M}$ is not diffeomorphic to $\mathbb{R}^4$), no set of four scalar fields can be invertible everywhere. Thus such scalar fields will not define a diffeomorphism between $\mathscr{M}$ and $\mathbb{R}^4$, and we cannot redefine the fields and the Lagrangian.  Hence, general relativity cannot be deparametrised in this way for non-trivial topologies.  A mathematical obstruction to deparametrising general relativity is also reported in reference~\cite{torre94}.\footnote{A further obstruction is that models with continuous symmetry are possible. In those cases, all sets of four scalar fields have zero Jacobian and consequently do not define a diffeomorphism even if $\mathscr{M}$ and $\mathbb{R}^4$ are diffeomorphic. However, the class of symmetrical models has zero measure and therefore it seems reasonable from a physical point of view to simply ignore such possibilities.}

One might also contemplate a more relaxed (and more complicated) strategy. Instead of finding a single set of four scalar fields, invertible everywhere on $\mathscr{M}$, one could try to extract a set of quadruples of scalar fields that define an atlas of the manifold.\footnote{More precisely, for each open region $\mathscr{U}_i$ in a set that covers $\mathscr{M}$, one must be able to assign {\em a priori\/} four scalar fields $x^\mu_i:\mathscr{U}_i\to\mathbb{R}$ (constructed from the unknown tensor fields $T$) which are guaranteed (except perhaps for a set of models with zero measure) to be invertible in $\mathscr{U}_i$ {\em for all\/} kinematically allowed models.} In non-pathological situations ({\em e.g.\/}, when continuous symmetries are absent) it is certainly always possible to extract such an atlas from the dynamical tensor fields for any {\em specific\/} model. However, it seems highly unlikely that any specific choice of atlas will work for all models with a specified topology, even if pathological cases ({\em e.g.\/}, models with continuous symmetries) are kinematically excluded.  Furthermore, such an atlas would not work for different topologies.

Thus, a feature that seems to distinguish general relativity from other field theories is not the general invariance of its dynamical equations, but its non-deparametrisability.

%-----------------------------------------------------------------------
\subsubsection{Dynamical isolation of the coordinate scaffolding}
\label{subsubsec:cr}
%-----------------------------------------------------------------------

As we already mentioned in section~\ref{subsec:coordinates}, in Newtonian mechanics and special relativity one can adopt a non-operational view of the coordinates $x^\mu$, in which these are regarded as mathematical parameters without any empirical content.  With this interpretation, in order to extract observables from such theories one must go through the same procedure as in general relativity.  However, in contrast to general relativity (and every other generally invariant theory), these theories also allow for an interpretation where the coordinates are  regarded as readings on some physical scaffolding.

Within such an interpretation, one sometimes also requires that there is a dynamical isolation between the coordinate scaffolding and the phenomena under investigation, so the scaffolding and the phenomena do not in some sense influence each others' behaviour.  For example, when one studies the propagation of a wave in a special relativistic context, the clocks and rulers used to define times and locations are assumed not to be affected by the presence of the wave itself.  Similarly, the details of the scaffolding do not enter in the wave equation, so the wave is not affected by the scaffolding.  Of course, this requires the existence of ``fiducial physical objects'', whose macroscopic behaviour is dynamically trivial in the sense that those properties relevant for defining a spatio-temporal coordinate grid, are not affected by the phenomena one wishes to observe.  For example, a coordinate scaffolding might ``heat up'' slightly from exposure to the wave.  However, such a difference is not necessarily relevant for the coordinate scaffolding's role as defining a spatio-temporal coordinate grid.

We can make these ideas more precise, and understand at the same time what it is that makes such a construction possible in pre-general relativistic theories, but not in general relativity.  For concreteness, imagine that one wants to study the propagation of a massless scalar field $\phi$ in Minkowski spacetime, which obeys
\begin{equation}
\frac{\partial}{\partial x^\mu}\left(\sqrt{-\eta}\,
\eta^{\mu\nu}\,\frac{\partial\phi}{\partial x^\nu}\right)=0
\label{kgdev}
\end{equation}
in coordinates devoid of any physical meaning ($\eta$ denotes the determinant of $\eta_{\mu\nu}$).  In addition to equation~(\ref{kgdev}), there will be some other differential equations describing the dynamics of the scaffolding, that we leave unspecified.  If we now assume that the
details of the coordinate scaffolding do not influence the dynamical
behaviour of the field $\phi$, then by identifying four coordinate fields $q^\mu$ on the scaffolding, equation~(\ref{kgdev}) becomes
\begin{equation}
\frac{\partial}{\partial q^\mu}\left(\sqrt{-\tilde{\eta}}\,
\tilde{\eta}^{\mu\nu}\,\frac{\partial\tilde{\phi}}{\partial
q^\nu}\right)=0\;,
\label{kgq}
\end{equation}
where
\begin{equation}
\tilde{\eta}^{\mu\nu}=\frac{\partial q^\mu}{\partial
x^\rho}\frac{\partial q^\nu}{\partial x^\sigma}\,\eta^{\rho\sigma}\;,
\label{tildeeta}
\end{equation}
$\tilde{\eta}$ is the determinant of $\tilde{\eta}_{\mu\nu}$, and $\tilde{\phi}$ is constructed as explained in section~\ref{subsubsec:scalars}.  In general, the field $\phi$ will affect the properties of the scaffolding, and in particular the functions $q^\mu(x)$ and so also $\tilde{\eta}^{\mu\nu}$; therefore~(\ref{kgq}), regarded as an equation for $\tilde{\phi}$, will be very messy and the dynamics of $\tilde{\phi}$ will not appear to have a well-defined structure.  However, one may limit oneself to use only the $q^\mu$ for which $\tilde{\eta}^{\mu\nu}$ does not depend of $\tilde{\phi}$ (provided that they exist), in which case no such problems would arise.  Equations~(\ref{kgdev}) and~(\ref{kgq}) will then be structurally identical, differing only by trivial relabeling.  One can therefore interpret the original $x^\mu$ as readings on a dynamically isolated coordinate scaffolding. 

Clearly, the equations describing the phenomenon of interest might, in general, contain not only the metric, but also other geometrical quantities such as the connection, and perhaps even non-geometrical ones.  The condition for being able to interpret the $x^\mu$ as readings on a dynamically isolated scaffolding is then that all these quantities, when expressed in terms of the $q^\mu$ as in equation~(\ref{tildeeta}), do not depend on the phenomenon itself.  This feature is by no means trivial: It is not guaranteed by any physical law, but requires careful design of the material scaffolding, which in particular must be shielded from the process under investigation.  As far as classical non-gravitational physics is concerned, it is in principle possible to keep the mutual influence arbitrarily small, but in practice only approximate isolation can be guaranteed.  Because of this, it is hard to believe that any theory that postulates a scaffolding given {\em a priori\/} could be fundamental.  This conclusion can be strengthened by considerations about the role and nature of the scaffolding within such theories~\cite{einstein-auto, brown}.

The idea of an isolated scaffolding is, however, untenable in general relativity.  Let us take again as an illustrative example the case of a massless scalar field, which in a generic curved spacetime obeys
\begin{equation}
\frac{\partial}{\partial x^\mu}\left(\sqrt{-\g}\,
\g^{\mu\nu}\,\frac{\partial\phi}{\partial x^\nu}\right)=0
\label{kgcurveddev}
\end{equation}
in coordinates without any physical meaning ($\g$ is now the determinant of $\g_{\mu\nu}$).  Introducing physical coordinates $q^\mu$, equation~(\ref{kgcurveddev}) becomes
\begin{equation}
\frac{\partial}{\partial q^\mu}\left(\sqrt{-\tilde{\g}}\,
\tilde{\g}^{\mu\nu}\,\frac{\partial\tilde{\phi}}{\partial
q^\nu}\right)=0\;,
\label{kgcurvedq}
\end{equation}
where
\begin{equation}
\tilde{\g}^{\mu\nu}=\frac{\partial q^\mu}{\partial
x^\rho}\frac{\partial q^\nu}{\partial x^\sigma}\,\g^{\rho\sigma}\;,
\label{tildeg}
\end{equation}
and $\tilde{\g}$ is the determinant of $\tilde{\g}_{\mu\nu}$.  Again, the condition of isolation of the scaffolding from the physical phenomenon under study is that $\tilde{\g}^{\mu\nu}$ remains the same for different solutions $\phi$.  (Note that by Einstein's equivalence principle the gravitational field $\g_{\mu\nu}$ must be present in the equations of motion, regardless which system one is studying.)  However, this cannot be the case in a general relativistic context, where $\phi$ itself influences the metric through Einstein's equations.  Hence, since gravity couples to everything, and everything influences the gravitational field, one cannot assume that the scaffolding is totally shielded from $\phi$.  Even if one could realise a direct shielding, an indirect one is always unavoidable, through the metrical field.\footnote{We tend to forget about this conceptually important point because, under ordinary conditions, the gravitational back-reaction of the observed phenomenon on the laboratory is very small, so one can assume, for all practical purposes (but not for issues of principle), that the laboratory is a fiducial physical object of the type described above.}  This is a more precise formulation of the idea that the behaviour of clocks and rods cannot be postulated {\em a priori\/} in general relativity, contrary to what happens in the analogous experiments performed within a special relativistic context.  

Note that the basic reasons why a coordinate scaffolding {\em a priori\/} does not exist in general relativity, are that every physical system affects $\g_{\mu\nu}$, and that no shielding from $\g_{\mu\nu}$ is possible.  This should not be confused with a different remark, that since $\g_{\mu\nu}$ can be different in various spacetime models, there is no universal prescription for the design of a scaffolding.  

We shall return to the intriguing interplay between the interpretation of coordinates, general invariance, and the relativity principle in section~\ref{subsec:general'}.

%-----------------------------------------------------------------------
\subsection{Invertibility}
\label{subsec:invertprob}
%-----------------------------------------------------------------------

An essential assumption for the viability of the construction of local observables outlined in section~\ref{subsec:localobservables} is that the function $q:\mathscr{U}\to {\cal Q}$ be invertible. But this is of course not guaranteed by any physical law.

If continuous symmetries (of both the gravitational and non-gravitational fields) are present, invertibility fails for {\em any\/} four scalar fields constructed out of the model, because the symmetry implies that the rank of the matrix with elements $\partial q^\mu/\partial x^\nu$ is strictly less than four.  These situations are, of course, pathological idealizations,\footnote{They can be reduced to normality just by adding non-trivial configurations of test fields, and then using these as the $q^\mu$.} but it is not entirely satisfactory that one cannot construct the above local observables in such cases.

Also, if the domain $\mathscr{U}$ of the four scalar fields is not appropriately restricted then the inverse image of some element in $\cal Q$ can contain more than one point of $\mathscr{M}$.  This means that the observables, \eg, $\tilde{s}(q)$, will be multivalued in general.  How many values they take, depends on how many elements of $\mathscr{M}$ are mapped in the same element of $\cal Q$ by the function $q$, which in turn is determined by the specific configuration $q^\mu(x)$.  In particular, this circumstance can become unavoidable for topological reasons (for example, a continuous function on the two-sphere that takes values in $\mathbb{R}^2$ cannot be invertible everywhere).  Hence, the observable $\tilde{s}$ is mathematically well-defined only locally, and what ``locally'' means can be specified only after one has a spacetime model.  This is, of course, not too problematic for classical general relativity, but when it comes to formulating a possible quantum theory of gravity, no spacetime model is specified, and the observables introduced in section~\ref{subsec:localobservables} are ill-defined and therefore not suitable for being turned into operators.

It is worth noticing that the relational observables can be understood as a particular form of gauge fixing.  Assuming invertibility, it is always possible to choose the particular gauge in which $q^\mu=x^\mu$ by a suitable spacetime diffeomorphism defined by the coordinate fields $q^\mu$ themselves.

%--------------------------------
\section{The space of point-coincidences}
\label{sec:events}
\setcounter{equation}{0}
%--------------------------------

We now outline a way of constructing local observables which does not suffer from the problem of invertibility pointed out in section~\ref{subsec:invertprob}.  The root of the problem lies in the fact that some scalar fields (\ie, the four scalar fields $q^\mu$), are selected to play a special role.  In the following we shall treat all dynamical degrees of freedom ``democratically'', so that no scalar fields will play any privileged role.

%------------------------------------------------------------------------
\subsection{An analogy: The parametrised curve}
\label{subsec:curve}
%------------------------------------------------------------------------

Consider a smooth curve $\mathscr{C}$ in $\mathbb{R}^n$.  It is natural to characterise $\mathscr{C}$ as the set of all ordered $n$-tuples $(\xi^1,\ldots,\xi^n)\in\mathbb{R}^n$ that satisfy a suitable system of equations $f_i(\xi^1,\ldots,\xi^n)=0$, with $i$ running from 1 to $n-1$.  For many purposes, it is however convenient to parametrise a portion of the curve by a label $\lambda\in\Lambda$, where $\Lambda$ is an open interval in $\mathbb{R}$.  This is possible because $\mathscr{C}$ is a differentiable manifold, so one can introduce charts on it.  We can thus describe, locally, the curve also through a mapping $\mbox{\boldmath $\xi$}:=(\xi^1,\ldots,\xi^n):\Lambda\to\mathbb{R}^n$ such that $f_i(\xi^1(\lambda),\ldots,\xi^n(\lambda))\equiv 0$ identically.  Of course, this parametrisation is not intrinsic to the curve and is totally arbitrary:  We are completely free to choose the one that better suits our needs, and the properties of the curve will not be affected by our choice.  More formally, we can say that such properties are invariant under reparametrisation.  In particular, if the map $\mbox{\boldmath $\xi$}$ were not defined directly, but only as the solution to some differential equations, such differential equations would exhibit invariance under an arbitrary differentiable change $\lambda\mapsto f(\lambda)$ of the parameter.

Indeed, this is the case if the curve can be described by a Lagrangian $L(\mbox{\boldmath $\xi$},{\rm d}\mbox{\boldmath $\xi$}/{\rm d}\lambda)$ such that $L(\mbox{\boldmath $\xi$},\epsilon\,{\rm d}\mbox{\boldmath $\xi$}/{\rm d}\lambda)=\epsilon L(\mbox{\boldmath $\xi$},{\rm d} \mbox{\boldmath $\xi$}/{\rm d}\lambda), \forall\epsilon>0$.  We are then dealing with $n$ scalar fields $\xi^1,\ldots,\xi^n$ defined on a ``parameter manifold'' where $\lambda$ is a local coordinate.  Since the theory is reparametrisation-invariant, if $\mbox{\boldmath $\xi$}(\lambda)$ is a solution then $\mbox{\boldmath $\xi$}(f(\lambda))$ is a solution, where $f$ is strictly monotonic.  Thus, there is an equivalence class of solutions, corresponding to different gauge choices, consistent with the same initial data (or boundary conditions).  Nevertheless, all members of a specific equivalence class describe the same curve in $\mathbb{R}^{n}$.  Mathematically speaking, the image $\mbox{\boldmath $\xi$}(\Lambda)$ does not depend on which member we pick.  On the other hand, the functional relation between the values of the scalar fields $\xi^1,\ldots,\xi^n$ and the value of the parameter $\lambda$, of course, varies from member to member.  Thus the gauge-independent data are contained in the image $\mbox{\boldmath $\xi$}(\Lambda)$ and not in the function $\mbox{\boldmath $\xi$}:\Lambda\to \mathbb{R}^{n}$.

%--------------------------------
\subsection{Point-coincidences}
\label{subsec:pc}
%--------------------------------

Events are usually taken as a primitive notion in relativity.  Basically, an event corresponds to some kind of ``happening'' --- for example, a collision between particles, or the emission of a flash of light.  Of course, this involves idealisations of the same type of those that lead to such ideas as ``point particles'' or ``light rays''.

Given the existence of physical quantities, whose values one can somehow measure, one can also adopt a more refined notion of an event as a ``point-coincidence'', expressed by the concomitant values of different quantities~\cite{kretschmann, einstein-1916}.  The precise nature of such quantities is irrelevant.  They could correspond to readings on some kind of coordinate scaffolding, as one usually does in special relativity, where it is the readings on a clock and on a grid of mutually orthogonal rigid rulers that define an event.  Or they could even be quantities upon which the experimentalist has no direct control, as it happens in cosmology, where events are defined by inhomogeneities in the large scale distribution of matter together with, {\em e.g.\/}, a value of the redshift.

It is an experimental fact that one need not know the values of all these quantities in order to identify uniquely a point-coincidence.  Indeed, locally the values of four of them are enough for this purpose.  And the choice of these four quantities is completely arbitrary, with the only caveat that one must avoid degenerate or pathological situations, so that one could tell one point-coincidence from another because the values of at least one of these quantities differ in the two cases.  Hence, the set of all point-coincidences possesses generically a natural manifold structure, obtained by interpreting the readings as coordinates in suitable charts (see, {\em e.g.\/}, reference~\cite{mtw}, pp.~5--13).  Hereafter, we shall denote such a set by $\mathscr{E}$.

%--------------------------------
\subsection{Defining the space of point-coincidences}
\label{subsec:pargrav}
%--------------------------------

Given that general relativity is a reparametrisation-invariant theory (\ie, it is generally invariant as defined in section~\ref{subsec:general}), it is natural to suspect that the gauge-invariant data in general relativity are contained in the image of the ``functions'' on the manifold, rather than in the functions themselves.  More specifically, suppose that $(\mathscr{M},T)$ is a model of spacetime.  Of course, one cannot just use the image of the tensorial functions on $\mathscr{M}$, because these are neither real-valued nor diffeomorphism-invariant, in general.  Also, one cannot use the image of their coordinate representations, since these are not invariant under coordinate transformations.  However, scalars are proper functions on the manifold $\mathscr{M}$, and their values {\em are\/} invariant under diffeomorphisms.  Suppose therefore that we can, from a given model $(\mathscr{M},T)$, construct a new one $(\mathscr{M},\Phi^1,\ldots,\Phi^N)$, where $\Phi^1,\ldots,\Phi^N$ are scalar fields which completely characterise the model, at least empirically.\footnote{As we already pointed out in section~\ref{subsec:localobservables} in relation to the physical coordinates $q^\mu$, such scalars need not be fundamental fields; they could well correspond to phenomenological properties.  In section~\ref{subsec:constructscalars} we consider some examples.}  An event corresponds to a joint reading of the values of all these scalars (a point-coincidence).  Hence, considering the map $\mbox{\boldmath $\Phi$}:=(\Phi^1,\ldots,\Phi^N):\mathscr{M}\to \mathbb{R}^N$, we can define the {\em space of point-coincidences\/}\footnote{As we shall see soon, the more appealing terminology of {\em manifold of point-coincidences\/} would not be mathematically accurate, because in pathological situations $\mbox{\boldmath $\Phi$}(\mathscr{M})$ is not a manifold.} $\mathscr{E}:=\mbox{\boldmath $\Phi$}(\mathscr{M}) \subset\mathbb{R}^N$.  This is a set of ordered $N$-tuples of real numbers, so it seems to have little to do with the four-dimensional spacetime of general relativity.  However, since $\mathscr{M}$ is four-dimensional, the rank of the matrix with components $\partial\Phi^H/\partial x^\mu$ (the index $H$ runs from 1 to $N$) cannot be greater than 4, so $\mathscr{E}$ is also four-dimensional, excluding pathological situations.

One can rephrase this conclusion saying that the scalars $\Phi^1,\ldots,\Phi^N$ are not independent, so in any specific model their values will be constrained by conditions of the type\footnote{The functions $F_I$ are, however, not unique.  They can be changed at will, provided that the subset of $\mathbb{R}^N$ defined by the simultaneous validity of equation~(\ref{FA}) for all values of $I$ remains the same.}
\begin{equation}
F_I(\Phi^1,...,\Phi^N)=0\;,
\label{FA}
\end{equation}
where $I$ runs from 1 to an integer number $M$ smaller than $N$.  In generic situations, these conditions define an $(N-M)$-dimensional submanifold in $\mathbb{R}^N$ by the implicit function theorem, with $M=N-4$.  It is obvious that $\mathscr{E}$ is invariant under diffeomorphisms of $\mathscr{M}$, so it contains all the local gauge-independent data.  Instead of using parameters/coordinates to mathematically characterise the totality of point-coincidences, one can instead characterise it implicitly through equations~(\ref{FA}).  Thereby, the use of coordinates is completely eliminated and one is left only with structure that is empirically accessible (at least in principle): The point-coincidences.

If, for all pairs of distinct points $p_1,p_2\in \mathscr{M}$, at least one of the functions $\Phi^1,\ldots,\Phi^N$ takes different values on $p_1$ and $p_2$, then the space of point-coincidences $\mathscr{E}$ is also a manifold, with the same dimension as $\mathscr{M}$.  However, if the values of the set of scalars are the same at two points $p_1$ and $p_2$ on the manifold $\mathscr{M}$, so that $\mbox{\boldmath $\Phi$}(p_1)=\mbox{\boldmath $\Phi$}(p_2)$, the points $p_1$ and $p_2$ correspond to only one point of $\mathscr{E}$.\footnote{Identifying events with the same properties as one and the same event is a straightforward application of Leibniz's principle of the identity of indiscernibles.}  Therefore, in the presence of continuous symmetries the dimensionality of the space of point-coincidences is lower than that of $\mathscr{M}$.\footnote{In the extreme case of Minkowski spacetime $\mathscr{E}$ contains just one point.}  However, all these situations are pathological, and this problem is far less serious than the one described in section~\ref{subsec:invertprob}.

The analogy between the construction of the space of point-coincidences and the pa\-ra\-me\-trised curve of section~\ref{subsec:curve} should be evident.  The unphysical coordinates $x^\mu$ used in general relativity correspond to the arbitrary parameter $\lambda$.  Also, the scalars $\Phi^1,\ldots,\Phi^N$ and the functions $F_1,\ldots,F_M$ that constrain them, correspond to $\xi^1,\ldots,\xi^n$ and $f_1,\ldots,f_{n-1}$.  Finally, $\mathscr{E}$ corresponds to $\mathscr{C}$.  It is amusing that almost everything that has been said in the philosophical debate about spacetime in general relativity can be repeated for the parametrised curve.  One notable exception, however, is the unphysical parameter manifold $\mathscr{M}$, which has no counterpart ($\Lambda$ does not correspond to $\mathscr{M}$, but rather to an open set in $\mathbb{R}^4$ --- a  coordinate domain).  In order to make the analogy complete, one should also introduce a parameter manifold on which $\lambda$ can be regarded as a local coordinate.  Of course, such a manifold would have a totally fictive character, in contrast to $\mathscr{C}$, and would be useless and rather confusing.  One may argue that this is the case also for $\mathscr{M}$ in general relativity.

%--------------------------------
\subsection{Constructing a complete set of scalars: Examples}
\label{subsec:constructscalars}
%--------------------------------

In Nature there is a multitude of different physical fields that could be used in order to construct the space of point-coincidences.  There are the various fermionic fields describing the three families of leptons pairs and the six quarks.  Then we have the bosonic fields describing the different interactions---the electromagnetic, weak, and strong forces.  Finally, there is the gravitational field.  Since most of these quantities are not scalars,  they cannot be used directly in order to define the space of point-coincidences $\mathscr{E}$.  One could then try to construct a complete set of scalars out of these fields and so construct $\mathscr{E}$.

However, this would not be terribly interesting, because most of these fields are not physically meaningful if considered classically. On the other hand, treating them as quantum fields would force one to generalize the concept of point-coincidence to the quantum domain, and it is not clear how to do that.\footnote{For example, in standard interpretations of quantum theory it is commonplace to assume that fields do not have precise objective properties, independent of measurement (although formulations do exist where they have precise properties at all times).  Since the construction of $\mathscr{E}$ critically depends on the specification of such precise properties, one has to conclude that $\mathscr{E}$ is ill-defined in the quantum domain, or at least becomes ``fuzzy'' in some sense according to the standard interpretations.}  Therefore, we restrict our attention to the macroscopic domain where classical physics applies.  Instead of fundamental physical fields we then deal with phenomenological fields like, \eg, the currents $j^\mu_a$ corresponding to the various chemical elements (the index $a$ runs from 1 to the  number of elements in the periodic table).

Consider then, as a concrete example, the following model of spacetime: $(\mathscr{M},\g_{\mu\nu},F_{\mu\nu},j^\mu_a)$.  Within it, we can construct a profusion of different scalar fields.  From the phenomenological chemical currents and the metric, one can construct the scalars $\g_{\mu\nu}\,j^\mu_a\, j^\nu_b$.  One may also construct the $14$ zeroth-order curvature invariants of the metric, and the two electromagnetic invariants $F_{\mu\nu}F^{\mu\nu}$ and $\epsilon_{\mu\nu\rho\sigma} F^{\mu\nu}F^{\rho\sigma}$, where $\epsilon_{\mu\nu\rho\sigma}$ is the Levi-Civita tensor.  And much more.

Denote this collection of scalar fields $\{\phi^A\}$.  Except in pathological situations, the matrix with components $\partial \phi^A/\partial x^\mu$ has rank 4 everywhere on $\mathscr{M}$.  We then construct other scalars as follows:
\begin{equation}
j^A_a=\frac{\partial \phi^A}{\partial x^\mu}\,j^\mu_a\;;\qquad
\g^{AB}=\frac{\partial \phi^A}{\partial x^\mu}\frac{\partial
\phi^B}{\partial x^\nu}\,\g^{\mu\nu}\;;\qquad
F^{AB}=\frac{\partial \phi^A}{\partial x^\mu}\frac{\partial
\phi^B}{\partial x^\nu}\,F^{\mu\nu}\;.
\label{FAB}
\end{equation}
Since the matrix  with components $\partial \phi^A/\partial x^\mu$ has rank 4, all the gauge-invariant information in the model is contained in the collection of all the scalars $\{\phi^A,j^A_a,\g^{AB},F^{AB}\}$.  We can then choose them as the fields $\Phi^1,\ldots,\Phi^N$.

At first it might seem typical that two or more events in the universe would share the same properties, \ie, that the values of the scalar fields $\Phi^1,\ldots,\Phi^N$ are the same for two points on $\mathscr{M}$. In such a case these points would be identified as one point-coincidence in $\mathscr{E}$.  One  might therefore doubt that events can really be characterised completely in terms of the local properties. But given the number of independent physical fields and the profusion of independent scalars that can be constructed therefrom, this case is in fact highly unlikely and not at all typical.  More precisely, the space of point-coincidences $\mathscr{E}$ is a four-dimensional subset in an $N$-dimensional space $\mathbb{R}^N$ where $N$ is much larger than $4$.  The probability that such a subset crosses itself is zero with respect to any reasonable measure on the space of four-dimensional subsets.  This conclusion can presumably be strengthened by considering the dynamical equations for the fields, and noting that solutions cannot intersect themselves in phase space.

%-----------------------------------------------------------------------
\section{Ontology of spacetime}
\label{sec:ontology}
\setcounter{equation}{0}
%-----------------------------------------------------------------------

One intuitive way of thinking about spacetime is that it constitutes a ``container'' for the histories of fields and particles.  Sometimes, this idea is blended with the non-operational interpretation of the coordinates, so this spacetime container is mathematically represented by the manifold $\mathscr{M}$, whose points cannot be identified operationally, because of the hole argument.  Adopting this view is logically possible and technically harmless, but it should be clear that the theory does not require it.  It fact, any notion of a spacetime behind the readings is, strictly speaking, metaphysical --- the predictions of the theory are only about the relations between physical readings and other physical readings.  In this paper we have seen a more sober ontology of spacetime emerging.  The space of point-coincidences $\mathscr{E}$ is a four-dimensional manifold in generic situations, contains all local gauge-independent data, and its elements represent physical events (which are, without exception, characterised by concrete properties).  It is therefore natural to identify $\mathscr{E}$ with spacetime itself, consistently with the intuitive idea that spacetime is the collection of all events.  (This view should not be confused with another one, according to which the fields somehow individuate independently existing points of $\mathscr{M}$~\cite{stachel}.)

Identifying spacetime with the space of point-coincidences, it follows that spacetime is conceptually defined by observable properties of the physical and geometrical fields (the point-coincidences).  It is a collection of properties, not a container physical objects are {\em in\/}.  This is probably the meaning of the remarks by Einstein, quoted in section~\ref{sec:intro}.  Indeed, without fields there is no spacetime --- now a trivial statement, since point-coincidences are defined only in terms of field values.  In particular, it makes no sense to think of a region of spacetime where there are no fields at all (no electromagnetic field, no scalar field, etc., and in particular no metric field).  That region would simply not exist at all~\cite{einstein}.

This view, in which spacetime acquires meaning only after field configurations are defined, is not in contradiction with the standard presentations, in which one first defines a manifold, then assigns tensor fields on it.  Indeed, physical quantities are mathematically represented by fields on a parameter manifold $\mathscr{M}$, not on the spacetime $\mathscr{E}$, which is defined only once a specific field configuration is given.  Consistently, field equations are also formulated on $\mathscr{M}$, not on $\mathscr{E}$.  A difficulty appears only if one tacitly identifies spacetime with $\mathscr{M}$.

What is, then, the role of the non-ontological parameter manifold $\mathscr{M}$?  Usually, one takes some features of dynamics as somehow ``induced'' by the topological and differentiable properties of a spacetime on which the physical quantities ``live''.  We suggest to adopt a shift in perspective.  Since all field equations are just partial differential equations in four independent parameters, the parameter space $\mathscr{M}$ naturally turns out to be a four-dimensional differentiable manifold.  Thus, dimensionality and differentiability of $\mathscr{M}$ represent the structure common to the dynamics of all fields (see reference~\cite{brown} for a convincing presentation of the idea that ``spacetime structure'' in general is actually rooted in physical dynamics).  This, however, does not give any ontological status to its points --- not anymore than the parameters $x^\mu$ have.  With some hindsight one could say that, ironically, it is most unfortunate that one has to write down field equations for the various fields, because in doing so one needs parameters (the $x^\mu$) which, although arbitrary, suggest that there is a manifold $\mathscr{M}$ given {\em a priori\/} out there.

A good summary of the situation is offered by the following comment by Einstein~\cite{jammer}:
\begin{quote}
{\em [...] the whole of physical reality could perhaps be represented as a field\footnote{Note that Einstein uses the singular ``field'' instead of our ``fields'', probably because of his belief in a unified field theory.} whose components depend on four space-time parameters.  If the laws of this field are in general covariant, that is, are not dependent on a particular choice of co\"ordinate system, then the introduction of an independent (absolute) space is no longer necessary.  That which constitutes the spatial character of reality is then simply the four-dimensionality of the field.  There is then no ``empty'' space, that is, there is no space without a field.\/}
\end{quote}

In fact, the container view of spacetime is {\em never\/} mandatory, even in pre-general relativistic spacetime theories.  For example, special relativity can be regarded as a theory with a distinct operationalistic character.  In it, physical fields are not expressed as functions of ``space'' and ``time'', but of actual measurements of distance and duration.  It is only when a background manifold is introduced, that the theory appears to possess unobservable, metaphysical elements; but this step is not a compulsory one, since at the very beginning the theory is formulated only in terms of readings on a coordinate scaffolding.  Interestingly, this is also the case for Newtonian dynamics.  Regarding the $x^\mu$ merely as coordinate readings, the theory is fully operational, since it is entirely formulated in terms of point-coincidences.  However, if one introduces Newton's absolute space --- a metaphysical concept, not necessary in order for the mathematics and the physical predictions of the theory to make sense ---, then the theory seems to possess unobservable features.  Thus, one should carefully distinguish between Newton's physics and his metaphysics.  In fact, one could strip the {\em Principia\/} of all the talk about absolute space, and still remain with a perfectly meaningful, although not fundamental (see section~\ref{subsubsec:cr}), physical theory.\footnote{In the field of investigation that is commonly placed under the heading of ``Mach's principle'', the Newtonian ideas of an absolute space (metaphysical) and of a standard of zero acceleration (physical) are often mixed together. Hence, although part of the critique is physically interesting (what determines inertial frames?  why are the compass of inertia and the compass of matter the same?), the arguments are blurred by attacks on Newton's absolute space, which need not play any role in Newtonian physics.}

%--------------------------------------------------------------
\section{Invariance {\em versus\/} relativity}
\label{sec:relativity}
\setcounter{equation}{0}
%------------------------------------------------------------------------

In section~\ref{sec:invariance}, Lorentz invariance and general invariance have been introduced in a similar way, as symmetry properties of field equations.  Lorentz invariance is commonly associated with the principle of special relativity, so it is natural to wonder whether some kind of relativity principle is also associated with general invariance.  In fact, this is often taken for granted in the literature about general relativity: Just as Lorentz invariance is a precise mathematical statement of the equivalence of all inertial frames of reference, so general invariance is assumed to guarantee that all frames, in arbitrary relative motion, are equivalent for the formulation of physical laws.  We now want to explore this issue in some depth.

%--------------------------------
\subsection{Lorentz invariance}
\label{subsec:lorentz'}
%--------------------------------

Let us start from Lorentz invariance, and try to understand in what sense it implies the validity of the principle of special relativity.  First of all, let us clarify the latter.  The main idea behind any relativity principle is the following one: A physical system (often referred to as the ``laboratory'') can be in one of several different configurations, that are physically distinguishable by making reference to the external world, but indistinguishable by means of internal experiments~\cite{brown-sypel}.  As it was noticed by Galileo, this is indeed the case if one considers mechanical experiments performed on two ships, one of which is at rest with respect to the shore, while the other is sailing at a constant velocity with respect to it. If it were possible to detect such a motion by means of other, non-mechanical, types of internal experiments, the relativity would be confined to a subset of physical phenomena, and we shall refer to such a situation, where ``internal indistinguishability'' holds only with respect to some kinds of phenomena, as a {\em weak\/} relativity principle.  On the contrary, if internal distinction is never possible, no matter what type of phenomena one uses, we shall speak of a {\em strong\/} relativity principle.  Hence, Galileo's relativity principle, which is restricted to mechanics, can be regarded as a weak one, whereas Einstein's relativity principle, which is supposed to refer to all types of phenomena, is strong.  Clearly, the validity of a weak relativity principle is a necessary condition for its strong version to hold.  Thus, in order to demonstrate that a strong relativity principle is not satisfied, it is enough to show that this is the case for a weak version of it.

Imagine, now, that one has some set of equations for a collection $\Psi$ of fields, formulated in terms of four parameters $x^\mu$.  Assume, further, that these equations are Lorentz-invariant, so from a given solution $\Psi(x)$ one can construct a new one, $\bar{\Psi}(x)=\Psi(\Lambda x)$, in the way described in section~\ref{subsec:lorentz}.  Mathematically, the function $\bar{\Psi}$ is obtained from $\Psi$ boosting it against the ``background'' given by the parameters $x^\mu$.

Although Lorentz invariance is present at a formal level, this situation does not necessarily correspond to the physical principle of special relativity.  Indeed, if the $x^\mu$ have no physical meaning (see section~\ref{subsec:coordinates}), the two solutions $\Psi(x)$ and $\bar{\Psi}(x)$ are indistinguishable not only internally, but also externally.  Since the $x^\mu$, by hypothesis, do not refer to anything empirical, the functions $x\mapsto\Psi(x)$ and $x\mapsto\bar{\Psi}(x)$ are not observable.  Observable quantities could be constructed by first choosing, between the fields present in $\Psi$, suitable coordinate fields $q^\mu$, then following the procedure outlined in section~\ref{sec:observables}.  However, in so doing the observable quantities will turn out to be the same for the two, mathematically distinct, configurations $\Psi(x)$ and $\bar{\Psi}(x)$, which contain the same physical point-coincidences.\footnote{Alternatively, one can say that the two models $(\mathscr{M},\Psi)$ and $(\mathscr{M},\bar{\Psi})$ are Leibniz-equivalent (see Appendix), so they correspond to the same physical situation.} Therefore, one of the conditions for having a physical relativity principle --- that of being able to distinguish, by making reference to the ``external world'', between different configurations of the ``laboratory'' --- is violated.

Of course, no such problem would arise if the variables $x^\mu$ were not unphysical parameters, but physical coordinates --- \ie, readings of some  suitably isolated scaffolding as discussed in section~\ref{subsubsec:cr}.  In this case, the two functions $\Psi(x)$ and $\bar{\Psi}(x)$ directly express observable point-coincidences between quantities in the ``laboratory'' (the values of $\Psi$ and of $\bar{\Psi}$) and in the ``external world'' (the values of the $x^\mu$).  These observables are different in the two situations, that are therefore distinguishable by ``external'' experiments, which use the coordinates $x^\mu$.  However, the ``internal'' point-coincidences remain the same in the two cases.  Together, these two facts guarantee that the principle of special relativity holds.

Note that the very possibility of writing equations for the quantities in $\Psi$ directly in terms of physical coordinates, without the need of introducing unphysical parameters, requires dynamical isolation between the coordinate scaffolding and $\Psi$.  Physically, this means that the laboratory does not back-react on the rest of the world (in particular, on the coordinate scaffolding), and is also not affected by it (see section~\ref{subsubsec:cr}).  This is an assumption that one usually takes for granted in the context of Newtonian mechanics and special relativity.

%--------------------------------
\subsection{General invariance}
\label{subsec:general'}
%--------------------------------

Consider now a generally invariant spacetime theory.  From any solution $T(x)$ we can construct a new one $\bar{T}(x)$ following the technique described in section~\ref{subsec:general} for the case of Einstein's equations.  This situation seems to closely mirror the case of Lorentz invariance just discussed, so it is natural to ask whether one can extract, from general invariance, a physical relativity principle.

For Lorentz invariance, it was possible to obtain a relativity principle regarding the $x^\mu$ as readings on some type of coordinate scaffolding.  This cannot be done for general invariance because, as we saw in section~\ref{subsec:hole-coord}, such an interpretation of the $x^\mu$ is untenable in that case.  The other possibility is to regard the $x^\mu$ as unphysical parameters, but then we run into the difficulty already met in section~\ref{subsec:lorentz'}: The solutions $T(x)$ and $\bar{T}(x)$ contain the same point-coincidences, hence they are physically indistinguishable and thus one of the conditions for having a relativity principle is violated.

There is still another possibility for associating a physical relativity principle with general invariance.  Let us suppose that the physical fields $T$ can be decomposed into two sets $\Upsilon$ and $\Psi$, such that if $(\Upsilon(x),\Psi(x))$ is a solution of the field equations, also $(\Upsilon(x),\bar{\Psi}(x))$ is, where $\bar{\Psi}(x)$ is constructed from $\Psi(x)$ using an arbitrary diffeomorphism, as discussed in section~\ref{subsec:general}.  If this is possible, the internal point-coincidences of $\Psi$ and $\bar{\Psi}$ are, of course, the same.  However, the point-coincidences of the total sets, $(\Upsilon,\Psi)$ and $(\Upsilon,\bar{\Psi})$, differ from each other.  Hence, regarding the fields that compose $\Psi$ as the ``laboratory'', and those that enter in $\Upsilon$ as the ``external world'', the conditions are fulfilled for having a relativity principle, at least in a weak form.

However, due to the arbitrariness of such a transformation, one could choose it in such a way that $(\Upsilon,\Psi)$ and $(\Upsilon,\bar{\Psi})$ satisfy the same initial value problem.  Then, if these two solutions are taken to correspond to different point-coincidences, as it should be the case in order to have a physical relativity principle, one would have that the same initial conditions can evolve into physically distinguishable configurations.\footnote{This does not happen in special relativity because a Lorentz transformation, unlike an arbitrary one, is ``rigid'', so one cannot find pairs of solutions $(\Upsilon,\Psi)$ and $(\Upsilon,\bar{\Psi})$ that are physically different and yet satisfy the same initial value problem.}  Hence, unless one is prepared to accept that a theory could be unable to predict point-coincidences in a unique way, one has to conclude that there is no relativity principle associated with general invariance.

One can also view this result from a different angle.  A sufficient condition for $(\Upsilon(x),\Psi(x))$ and $(\Upsilon(x),\bar\Psi(x))$ to be both solutions, and so for having point-coincidence under-determinism, is that the field equations for $\Upsilon$ and $\Psi$ be completely decoupled (\ie, the set of equations for $\Upsilon$ does not contain $\Psi$, and vice versa).  Interestingly, such a possibility is forbidden in general relativity by the equivalence principle, according to which the gravitational field couples to all other fields.

In the case of parametrised field theories the situation is similar.  Consider a number of distinct fields collectively referred to as $\Psi(x)$, and let the corresponding field equations be Lorentz invariant.  Then we parametrise the theory as outlined in section~\ref{subsubsec:discussion} to obtain generally invariant equations for the fields $x^\mu(\xi)$ and $\Psi(\xi)$.  In this theory the coordinates $x^\mu$ couple universally to all fields, so no set of field equations are completely decoupled from any other, and no contradiction with determinism can therefore be established.  It is also interesting to notice that Lorentz invariance as a relativity principle is not lost, although it is now somewhat hidden.  It is still the case that if $(x^\mu(\xi),\Psi(\xi))$ is a solution, then so is $(x^\mu(\xi),\varphi_\ast\Psi(\xi))$, for any diffeomorphism $\varphi$ that identifies a Poincar\'e transformation.  Therefore, generally invariant equations can contain hidden symmetries which are not directly connected to the invariance properties of the equations as such.

It should be obvious, from the discussion about Lorentz invariance, that physical principles of relativity are rather messy, since they involve hypotheses like decoupling and isolation (see section~\ref{subsubsec:cr}), which are definitely not likely to be fundamental.  On the other hand, the notion of mathematical invariance is a very clean one; but, as we have seen in the case of general invariance, it is not always associated with a physical principle of relativity.

%--------------------------------------------------------------
\section{Conclusions}
\label{sec:discussion}
\setcounter{equation}{0}
%--------------------------------------------------------------

In order to make physical predictions from a theory, one must supplement the formalism with an interpretation.  In the particular case of spacetime theories, the physical meaning of the invariance properties relies on the meaning of the coordinates $x^\mu$.  We have considered two possible interpretations: In the first one, the coordinates represent physical readings and thus have operational significance; in the second one they are mere mathematical parameters, so they are operationally meaningless.  Then, we have pointed out that in theories with a fixed spacetime structure, like Newtonian mechanics and special relativity, both of these interpretations are possible; whereas in generally invariant theories like general relativity, only the second one is viable.  This is tantamount to saying that the manifold $\mathscr{M}$ cannot represent something empirically accessible in general relativity.  In Einstein's own words~\cite{einstein-1916}:
\begin{quote}%
{\em [...] this requirement of general co-variance [...] takes away from space and time the last remnant of physical objectivity.\/}
\end{quote}
Indeed, if we take the manifold $\mathscr{M}$ to be the mathematical representation of spacetime, then general invariance (or general covariance, as Einstein calls it) forces us to accept that space and time have no empirical significance.

This conclusion leads one to wonder how spacetime and observable quantities can be described in general relativity.  As we have seen, the space of point-coincidences $\mathscr{E}$ is a natural representation of the totality of physical events (\ie, of spacetime), and at the same time contains all the observables of the theory.  In this representation the manifold $\mathscr{M}$ plays no empirical role, and everything that is physically relevant are the mutual relationships of the configurations of various fields.   This can be regarded as corresponding to some kind of relational ontology.

Identifying spacetime with the space $\mathscr{E}$ of point-coincidences offers a possibility for making precise some notions that one sometimes encounters in the literature about quantum gravity, such as ``fuzzy spacetime'' or ``fractal spacetime''.  Indeed, from the perspective here developed it is somewhat unnatural that the set $\mathscr{E}$ should behave as a four-dimensional smooth manifold everywhere and at every resolution.

%--------------------------------------------------------------
\section*{Acknowledgements}
%--------------------------------------------------------------

It is a pleasure to thank Julian Barbour, Harvey Brown, Bianca Dittrich, Lucien Hardy, Joe Henson, Brian Pitts, Oliver Pooley, Simon Saunders, Rafael Sorkin, Ward Struyve, and Antony Valentini for stimulating discussions and correspondence.

%--------------------------------------------------------------
\section*{Appendix: Leibniz equivalence}
\label{sec:leibniz}
\setcounter{equation}{0}
%--------------------------------------------------------------

When one formulates a spacetime theory in terms of an abstract parameter manifold $\mathscr{M}$ whose points are assumed to be devoid of any operational meaning, this manifold is sometimes regarded as the seat for ``potential events'', independent of whether such events are actually realised or not.  The role of $\mathscr{M}$ becomes then analogous to the one played by the absolute space in Newtonian mechanics, as a ``container'' for physical bodies (a set of ``potential positions'' for pointlike particles).

Such an interpretation of $\mathscr{M}$ suffers from the following difficulty.  Suppose that $(\mathscr{M},T)$ is a model of spacetime.  Consider another manifold $\mathscr{M}'$ diffeomorphic to $\mathscr{M}$ under a map $\varphi:\mathscr{M}\to \mathscr{M}'$.  The diffeomorphism $\varphi$ induces an application $\varphi_\ast$ that associates to the fields $T$ defined on $\mathscr{M}$ new corresponding fields $\varphi_\ast T=:T'$, defined on $\mathscr{M}'$. The spacetime model $(\mathscr{M}',T')$ so obtained is, in general, mathematically distinct from $(\mathscr{M},T)$.  However, they are physically indistinguishable, because the map $\varphi_\ast$ acts on {\em all\/} fields, thus leaving unaltered the results of any conceivable experiment (see also reference~\cite{wald}, p.~438).  Indeed, experimental results always correspond to suitable point-coincidences, which are the same in $(\mathscr{M},T)$ and $(\mathscr{M}',T')$.

In order to further clarify this important issue, let us consider the case $\mathscr{M}'=\mathscr{M}$.  Under a diffeomorphism $\varphi: \mathscr{M}\to \mathscr{M}$, one gets different functions $T$ and $T'$ on the same $\mathscr{M}$, with $T(p)\neq T'(p)$ in general, for $p\in \mathscr{M}$.  Hence, the association between manifold points and field values is, in general, scrambled.  In order to distinguish between the two models $(\mathscr{M},T)$ and $(\mathscr{M},T')$, one should be able to detect the difference between the values of $T$ and $T'$ at a given point $p\in \mathscr{M}$, so one should first identify the point $p$ by suitable measurements.  However, by assumption the points of the abstract manifold $\mathscr{M}$ lack operational significance by themselves, and whatever physical field one may try to use for this purpose is also affected by $\varphi$, so one cannot just ``read off'' points of $\mathscr{M}$ by the field values.\footnote{In other words, coordinates associated with readings on physical reference frames cannot be kept ``anchored'' to manifold points.  Of course, this would instead be possible for purely abstract, mathematical coordinates, but these are useless when it comes to operationally deciding between $(\mathscr{M},T)$ and $(\mathscr{M},T')$.}  Hence, there is no operational way to tell $(\mathscr{M},T)$ from $(\mathscr{M},T')$.  (Note, however, that if the points of $\mathscr{M}$ are instead assumed to have operational meaning, then the two models $(\mathscr{M},T)$ and $(\mathscr{M}, T')$ do not have the same empirical content.)

One is thus led to identify a physical situation not just with a single model $(\mathscr{M},T)$, but with the entire equivalence class $[(\mathscr{M},T)]$ of all such models, related to each other by arbitrary diffeomorphisms (see reference~\cite{he}, p.~56).  In the philosophical literature, the fact that two diffeomorphic models are physically indistinguishable is referred to as {\em Leibniz equivalence\/}~\cite{hole}.  One can regard Leibniz equivalence as expressing a gauge freedom of spacetime theories\footnote{All spacetime theories where $\mathscr{M}$ has no operational significance involve such a gauge freedom.  However, generally invariant theories are gauge theories even in a more strict, technical sense; that is, initial data do not uniquely specify the future evolution.} --- the points of the manifold $\mathscr{M}$ constitute excess baggage,\footnote{A manifold is, in the first place, a set; hence its points carry properties that serve to distinguish them from each other.} not encoded in the physical field configuration.  Of course, all this can be repeated for {\em space\/} in pre-relativistic theories as well, provided that the non-operational interpretation of the manifold points is adopted.  Indeed, Leibniz equivalence was first formulated in order to argue against Newtonian absolute space~\cite{barbour}.

%--------------------------------------------------------------
%
\end{document}